\documentclass[journal]{IEEEtran}
\usepackage{mathtools,amssymb,bm}
\usepackage{amsmath}
\usepackage{cite}
\usepackage{graphicx}
\usepackage{epstopdf}
\usepackage{cite}
\usepackage{color}
\usepackage{caption}
\usepackage[footnotesize]{subfigure}
\usepackage{hyperref}
\newtheorem{thm}{Theorem}[section]
\newtheorem{lem}[thm]{Lemma}
\newtheorem{prop}[thm]{Proposition}

\newtheorem{defn}{Definition}[section]

\begin{document}
	\title{Robustness of Two-Dimensional Line Spectral Estimation Against Spiky Noise   }
	\author{ Iman Valiulahi, Farzan Haddadi, and Arash Amini
}
	\maketitle
	\begin{abstract}
		The aim of two-dimensional line spectral estimation is to super-resolve the spectral point sources of signal from time samples. In many associated applications such as radar and sonar, due to cut-off and saturation regions in electronic devices, some of the number of samples are corrupted by spiky noise. To overcome this problem, we present a new convex program to simultaneously estimate spectral point sources and spiky noise in two dimension. To prove uniqueness of the solution, it is sufficient to show that a dual certificate exists. Construction of the dual certificate imposes a mild condition on the separation of the spectral point sources. Also, the number of spikes and detectable sparse sources are shown to be a logarithmic function of the number of time samples. Simulation results confirm  the conclusions of our general theory.
			 \end{abstract}
	\begin{IEEEkeywords}
		Two-dimensional, line Spectral estimation, total variation norm, continuous dictionary, convex optimization.
	\end{IEEEkeywords}
\section{Introduction}
Multiple-dimensional line spectral estimation (LSE) has received much attention in signal processing society in recent years. It is a fundamental concern of many applications such as Multiple-Input Multiple-Output (MIMO) radar \cite{heckel2017generalized}, super-resolution imaging \cite{huang2008three} and channel estimation in wireless communications \cite{gao2014super}. Specifically, in the MIMO radar, there exist multiple transmit antennas which emit probing signals to the target, a reflected version of these signals is then received by multiple receiver antennas. One can super-resolve the triples relative angle, Doppler shifts, and delay to estimate the location and velocities of the targets. In the LSE setting, a linear combination of $r$ sinusoidal with arbitrary complex amplitude should be observed. In practice, a subset of the observations is corrupted by spiky noise due to cut-off and saturation in  electronic devices. In this paper, we address the two-dimensional LSE problem in the presence of spiky noise.\par 
Parametric approaches in the LSE problem are based on dividing the observation space into signal subspace and noise subspace by singular value decomposition \cite{hua1992estimating}, \cite{hua1993pencil}. Although the computational complexity of these approaches is low, the sensitivity to additive perturbation is high. Also, they can not determine the location of spiky noise and require prior knowledge of number of spectral sources.

In another field of study, known as compressive sensing, an exact solution can be achieved in an underdetermined linear system by assuming that the signal of interest is sparse in a known discrete dictionary \cite{candes2006robust},\cite{donoho2006compressed}, here the Fourier domain. Therefore, $\ell_{1}$ minimization can be used to recover the support of the signal on DFT basis. However, in many practical applications such as radar and sonar, the spectral sources belong to a continuous dictionary. Thereby, mismatch between the actual sources and DFT basis is inevitable \cite{tropp2007signal} \cite{chi2011sensitivity}. In \cite{verzelen2012minimax}, it is shown that refining the grid exceedingly raises computational complexity. Further, $\ell_{1}$ minimization dose not achieve an exact solution in the ultrahigh-dimensional setting.
\par 
  The LSE model incorporates a two-dimensional spectral matrix $\bm{X} \in \mathbb{C}^{n_1\times n_2}$, given by \begin{align}\label{superposition}
X_{\bm{k}}=\sum_{i=1}^{r}d_i e^{j2\pi \bm{f}_{i}^T\bm{k}}, ~~~~~~~~ \bm{k} \in \mathcal{N},
 \end{align}
 where $\bm{d}$ is the complex amplitude vector, $d_i=|d_i|e^{j\phi_{i}},~\phi_{i} \sim \mathcal{U}(0,2\pi)$, $|d_i| \sim \delta_{0.5}+\mathcal{X}^{2}(1)$, $r$ is the degree of sparsity, $\bm{f}_i \in [0,1]^2$ and $\mathcal{N} = \{1,\cdots ,n\}\times\{1,\cdots,n\}$ is the two-dimensional integer square that indicates time indices.
 
 In \cite{candes2014towards}, Candes and Fernandez have proposed a non-parametric approach to super-resolve the inherent frequencies in (\ref{superposition}) for one and two dimensional cases. Their approach is based on Total-Variation (TV) norm minimization which is used to promote the sparsity of a continuous function. 
 It has been shown that a linear combination of the fourth power of Dirichlet kernel and its derivatives can be used to construct a valid certificate to this problem. This construction imposes a minimum separation $4/n$ and $4.76/n$ between the frequency sources in one and two dimensions, respectively. To achieve sharp bound on minimum separation in one-dimensional case, \cite{fernandez2016super} has constructed the dual certificate by three Dirichlet kernels with different cut-off frequencies. Also, \cite{valiulahi2017two} has extended this kernel to two-dimensional situation. The imposed minimum separations respectively are $2.52/n$ and $3.36/n$. In \cite{tang2013compressed}, one-dimensional signal is observed in a random subset of time instances. Tang et.al. have proved that under a mild assumptions on the minimum separation and with $\mathcal{O}(r \log r \log n)$ partial time samples, one can always find the random trigonometric polynomial that estimates the point sources. Their approach has been extended to two-dimensional case in \cite{chi2015compressive}.
 
 Additive noise is inevitable in most of the mentioned applications. Recently, a significant line of study is focused on  support stability of $\mathrm{TV}$ norm minimization when the measurements are corrupted by additive perturbation \cite{candes2013super}. Establishing a trade-off between  noise power and $\mathrm{TV}$ norm, known as BLASSO, is a common way to deal with such a problem \cite{poon2017multi}, \cite{de2012exact}. \cite{bhaskar2013atomic} made a precise comparison between robustness of optimization based methods and conventional approaches dealing with additive noise.  Due to sensor failure, another kind of corruption that may appear in the applications is spiky noise. Subspace decomposition approaches are not able to estimate the sources' frequencies when a subset of time samples is corrupted by spiky noise. Precisely, they are designed to overcome Gaussian-like noise and are relatively ill-positioned in front of spiky noise.
 
 \cite{fernandez2016demixing} has suggested a convex optimization program that incorporates the sparsity feature of spiky noise in the cost function in order to simultaneous estimation of both spectral sources and spiky noise in a one-dimensional situation. Although many related works in the literature have revolved around one dimension, but many applications such as super-resolution imaging and MIMO radar require multi-dimensional analysis. 
 
 In this paper, our main contribution is to construct a valid  dual certificate for two-dimensional case. Our certificate is  a  two-dimensional trigonometric low-pass polynomial that interpolates any sign pattern of the signal. Further, its coefficients belong to relative interior of sub-differential of $\ell_{1}$ norm. It is shown that if the number of the spectral sources and the spiky noise samples is restricted by a logarithmic function of the number of samples, our semidefint programing achieves an exact solution under mild condition on the separation between the spectral sources. \cite{chen2014robust} has proposed an innovative approach based on matrix completion that simultaneously super-resolves the spectral sources and detects spiky noise in two-dimensional case. It also proved that Enhanced Matrix Completion (EMaC) achieves an exact solution if the sample complexity exceeds  $\mathcal{O}(r^2 \log^3 n^2)$, under some incoherence conditions.
 
 The inherent infinite dimension of the LSE problem is apparently an enormous challenge. One can approximate the problem on a fine grid \cite{duval2017sparse} or solve the $\mathrm{TV}$ norm minimization directly by linear programing \cite{de2017exact}. \cite{candes2014towards} has converted the dual of $\mathrm{TV}$ norm minimization to linear matrix inequality (LMI) using positive trigonometric polynomial (PTP) theory \cite{dumitrescu2017positive}. The magnitude of the trigonometric polynomial in each subband of its frequency domain is controllable by the coefficients that are obtained from  PTP theory \cite{mishra2015spectral,valiulahi2017off,yang2018fast}. Finally, we present numerical simulations to justify our results. 
 
 The rest of the paper is organized as follows. The problem is formulated in Section \ref{sec.formulation}. Penalized $\mathrm{TV}$ norm minimization and our main theorem in two-dimensional case are presented in section \ref{TVnorm}. Construction of dual certificate and implementation of the dual problem are given in Sections \ref{refdual} and \ref{imdual}, respectively. Section \ref{exp} is devoted to numerical experiments. Finally, conclusions are discussed in Section \ref{Con}.
 
 \textbf{Notation}.
 Throughout the paper, scalars are denoted by lowercase letters, vectors by lowercase boldface letters, and matrices by uppercase boldface letters. The $i$th element of the vector $\bm{x}$ and the $\bm{k}=(k_1,k_2)$ element of the matrix $\bm{X}$ are given by $x_i$ and $X_{\bm{k}}$, respectively. $|\cdot|$  denotes cardinality of sets, absolute value for scalars and element-wise absolute value for vectors and matrices, also $\|\bm{z}\|_{\infty}=\underset{i}{\max}~|z_{i}|$. For a function $f$ and a matrix $\bm{A}$, $\|f\|_{\infty}$, $\|\bm{A}\|_{\infty}$, $\|\bm{A}\|$ and $\|\bm{A}\|_{1}$ are defined as $\underset{t}{\sup}|f(t)|$, $\underset{\|\bm{x}\|_{\infty}\le1}{\sup}\|\bm{Ax}\|_{\infty}=\underset{i}{\max}\sum_j|A_{i,j}|$, $\underset{\|\bm{z}\|_{2}\le1}{\sup}\|\bm{A}\bm{z}\|_{2}$ and $\underset{i,j}{\sum}|A(i,j)|$, respectively. $\mathrm{relint}(C)$ denotes relative interior of a set $C$. $\partial f(\cdot)(x)$ denotes sub-differential of  function $f$ at point $x$. $f^{i}(t)$ and $f^{i_1i_2}(\bm{t})$ denote $i$th derivate and $i_1,i_2$ partial derivatives of one-dimensional function $f(t)$ and two-dimensional function $f(\bm{t}:=(t_1,t_2))$, respectively. $(\cdot)^T$ and $(\cdot)^{*}$ show transpose and  hermitian of  a vector, respectively. $\mathrm{sgn}(\bm{x})$ denotes the element-wise sign of the vector $\bm{x}$. Also, $\mathrm{vec}(\bm{X})$ denotes the columns of $\bm{X}$ being stacked on top of 
 each other. The inner product between two functions $f$ and $g$ is defined as $\langle f,g \rangle:=\int f(t) g(t) dt$ and $\langle\cdot,\cdot\rangle$ denotes the real part of Frobenius product. $\otimes$ is Kronecker product and the adjoint of a linear operator $\bm{\mathcal{F}}$ is denoted by $\bm{\mathcal{F}}^*$.   
 \section{Problem Formulation}\label{sec.formulation}
 In the spectral domain the signal in (\ref{superposition}) is a linear combination of Dirac delta functions:
 \begin{align}\label{Driacmeasure}
 \bm{\mu}=\sum_{\bm{f}_i \in T}d_i \delta(\bm{f}-\bm{f}_i),
 \end{align}
 where $T$ is support of the signal and $\delta(\bm{f}-\bm{f}_i)$ denotes the Dirac delta function located in $\bm{f}_i$. The main goal in LSE is to recover the location and amplitude of each delta by finite time samples. As mentioned in the introduction, many practical applications such as radar and sonar suffer from spiky noise due to their electrical instruments, so we going to investigate the two-dimensional LSE when a subset of time samples is completely corrupted by spiky noise. Assume that spiky noise is added to the signal (\ref{superposition}) as:
 \begin{align}\label{noisyproblem}
 Y_{\bm{k}}=X_{\bm{k}}+Z_{\bm{k}}, ~~~~~\bm{k} \in \mathcal{N},
 \end{align}
  where $Z_{\bm{k}}$ is an element of the sparse two-dimensional noise matrix $\bm{Z} \in \mathbb{C}^{n\times n}$ with $s$ non-zero entries. The observation model can be written in the matrix form:
  \begin{align}
  \label{noisyoprator}
  \bm{Y}=\bm{\mathcal{F}}\bm{\mu}+\bm{Z},
  \end{align}  
  where $\bm{\mathcal{F}}(\cdot)$ is a linear operator that maps a continuous-indexed function in the frequency domain  to two-dimensional integer square $\mathcal{N}$ in time domain. The problem is to simultaneously estimate the spectral sources and the location of spiky noise form $\bm{Y}$.
  \section{Robust Total Variation Minimization }\label{TVnorm}
   The spectral sparsity is not sufficient to tackle this problem. In fact, if sources are located too close to each other,it would be impossible to resolve them \cite{candes2014towards}.
  \begin{defn} Let $\mathbb{T}^{2}$ be two-dimensional torus obtained by identifying the endpoints on $[0,1]^2$. For each set of points $T\subset\mathbb{T}^2$, the minimum separation is defined as:
  	\begin{align}
  	\label{eq6}
  	&\Delta(T):=\underset{\bm{t}_i,\bm{t}_j\in T,~ \bm{t}_i\neq \bm{t}_j}\inf~\|\bm{t}_i-\bm{t}_j \|_\infty\nonumber\\
  	&=\underset{i\neq j} \inf\max \{|t_{1i}-t_{1j}|,|t_{2i}-t_{2j}|\},
  	\end{align}
  	where $|t_{1i}-t_{1j}|,|t_{2i}-t_{2j}|$ denote wrap-around distances on the unit circle.
  \end{defn}

 $\ell_{1}$ minimization is not suitable for LSE problem, due to the discretization of spectral domain. Whereas,  $\mathrm{TV}$ norm can promote the sparsity of continuous functions and defined as $
  \|\bm{\nu}\|_{\mathrm{TV}}:=\sup_{\rho}\sum_{E\in\rho}|\bm{\nu}(E)|$.
 Indeed, $\mathrm{TV}$ norm maximizes the disjoint sum of positive measures $|\bm{\nu}(\cdot)|$ over all partitions $\rho$ of  square $[0,1]^2$. In the special case of (\ref{Driacmeasure}), $\|\bm{\mu}\|_{\mathrm{TV}}=\sum_{i=1}^{r}|c_{i}|$.
  
   Most inverse problems are solved by minimizing a cost function that promotes an inherent structure \cite{chandrasekaran2012convex}. This concept emerges in compressed sensing \cite{candes2006robust} and matrix completion \cite{candes2009exact}. The cost function may also include a specific penalty term to perform side tasks. For instance, \cite{chen2001atomic} has shown that penalizing $\ell_{1}$ norm  with an $\ell_2$ error term is an efficient way for denoising. The observation model (\ref{noisyoprator}) is sum of two sparse signals in different domains. \cite{fernandez2016demixing} balances the $\mathrm{TV}$ norm of spectral sources and the $\ell_{1}$ norm of spiky noise for one-dimensional case. We generalize this approach to two-dimensional case by introducing the following optimization problem:
 \begin{align}
 \label{primalproblem}
 \mathrm{P}_{\mathrm{TN}}:~~~\min_{\tilde{\bm{\mu}},\tilde{\bm{Z}}}~\|\tilde{\bm{\mu}}\|_{\mathrm{TV}}+\lambda\|\tilde{\bm{Z}}\|_{1} ~~\mathrm{subject\,\, to}~~ \bm{Y}=\bm{\mathcal{F}}\tilde{\bm{\mu}}+\tilde{\bm{Z}},\nonumber
 \end{align}
 where $\lambda > 0$ is a regularization parameter that makes a trade-off between $\mathrm{TV}$ norm of spectral spikes and $\ell_{1}$ norm of spiky noise. The first goal of this paper is to prove that $\mathrm{P}_{\mathrm{TN}}$ achieves exact recovery and the second is to provide a valid semidefinite programming to solve this problem.
   The following theorem states that the solution of $\mathrm{P}_{\mathrm{TN}}$ is exact, under specific conditions. 
 \begin{thm}\label{thm.main}
 	Let $J=\{1,\cdots,n\}\times\{1,\cdots,n\}$ be the set of observed entries in the matrix,
\begin{align}
\bm{Y}=\bm{\mathcal{F}}\bm{\mu}+\bm{Z},
\end{align} 
where each element of the noise matrix is independently non-zero with probability $\frac{s}{n^2}$ supported on $\Omega$ ($|\Omega|=s$). Also, the support of 
 	\begin{align}\label{Driacmeasure1}
 	\bm{\mu}=\sum_{\bm{f}_i \in T}d_i \delta(\bm{f}-\bm{f}_i),
 	\end{align}
 obeys
 	\begin{align}
 	\label{eq7}
 	\Delta(T) \geq \frac{3.36}{n-1},
 	\end{align}
 		where $\bm{f}_{i} \in [0,1]^2$ and $|T|=r$.	If $r+s\le n^2$,
 	\begin{align}
 	\label{boundonks}
 	&r \leq C_{r} \big(\log \frac{n^2}{\epsilon}\big)^{-2}n^2, &&s \leq C_{s} \big(\log \frac{n^2}{\epsilon}\big)^{-2}n^2,\nonumber\\
 	&n \geq 4 \times 10^3,&& \lambda=\frac{1}{\sqrt{n^2}},
 	\end{align}
 	then, the exact solution of $\mathrm{P}_{\mathrm{TN}}$ is   $(\bm{\mu},\bm{Z})$  with probability $1-\epsilon$ for any $ \epsilon >0$ and numerical constants $C_{r}$ and $C_{s}$.
 		Using Theorem \ref{thm.main} one can estimate the spectral sources and spiky noise samples up to logarithmic functions of the number of time samples under a mild condition on the separation of the spectral sources.
 	 	  \end{thm}
 \section{Construction of the Dual Certificate}\label{refdual}
The following proposition justifies that if there exists a low-pass trigonometric polynomial with coefficients in relative interior of sub-differential of $\ell_{1}$ norm which interpolates any signal sign pattern in $T$, then it is a sufficient certificate for Theorem \ref{thm.main}.
  \begin{prop}\label{proposition1}
 	If the conditions of Theorem \ref{thm.main} hold,
 	for any sign patterns  $\bm{h} \in \mathbb{C}^{|T|}$ and $\bm{r} \in \mathbb{C}^{|\Omega|}$ such that $|h_{i}|=1$ and $|r_{l}|=1$, for all $i$ and $l$, then there exists two-dimensional low-pass trigonometric polynomial  
 		\begin{align}
	\label{dualcerti}
	\bm{\mathcal{F}^{*}\bm{C}}=Q(\bm{f})=\sum_{\bm{k}\in J}C_{\bm{k}} e^{-j2\pi \bm{f}^T \bm{k}},
\end{align}
such that
	\begin{align}
&Q(\bm{f}_i)=h_{i},  &&\forall \bm{f}_i \in T\label{con1},\\
&|Q(\bm{f})| < 1,  &&\forall \bm{f} \notin T,\label{con2}\\
&\frac{C_{\bm{k}_{l}}}{\lambda} =r_{l}, && \forall \bm{k}_{l} \in \Omega\label{con3},\\
&|C_{\bm{k}}|<\lambda,&&\forall \bm{k}\notin \label{con4}\Omega,
 \end{align}
 where $\bm{f}_{i}=(f_{1i},f_{2i})$ and $\bm{k}_{l}=(k_{1l},k_{2l})$.
\end{prop}
(\ref{con1}) and (\ref{con2}) state that $Q(\bm{f}) \in \mathrm{relint}(\partial(\|\cdot\|_{\mathrm{TV}}(\bm{\mu})))$, so for any  measure $\hat{\bm{\mu}} $ 
 	\begin{align}
\|\bm{\mu}+\bm{\hat{\mu}}\|_{\mathrm{TV}} \geq \|\bm{\mu}\|_{\mathrm{TV}}+ \langle Q,\hat{\bm{\mu}}\rangle.
 \end{align}
  Similarly, it can be deduced from  (\ref{con3}) and (\ref{con4}) that $\frac{\bm{C}}{\lambda} \in \mathrm{relint}(\partial(\|\cdot\|_{1}(\bm{\mu}))) $, so for any  $\hat{\bm{Z}}$ we have
   	\begin{align}
 \|\bm{Z}+\hat{\bm{Z}}\|_{1} \geq \|\bm{Z}\|_{1}+ \langle \frac{\bm{C}}{\lambda},\hat{\bm{Z}}\rangle.
 \end{align}
 Let $\bar{\bm{\mu}}=\bm{\mu}+\bm{\hat{\mu}}$ and 
 $\bar{\bm{Z}}=\bm{Z}+\hat{\bm{Z}}$ as a feasible point such that $ \bm{Y}=\bm{\mathcal{F}}_{2D}\bar{\bm{\mu}}+\bar{\bm{Z}}$:
 \begin{align}
& \|\bar{\bm{\mu}}\|_{\mathrm{TV}}+\lambda\|\bar{\bm{Z}}\|_{1} \geq \|\bm{\mu}\|_{\mathrm{TV}}+\lambda\|\bm{Z}\|_{1}+\langle Q,\bar{\bm{\mu}}-\bm{\mu} \rangle\nonumber\\
 &+\lambda\langle \frac{\bm{C}}{\lambda},\bar{\bm{Z}}-\bm{Z}\rangle_{F}\geq \|\bm{\mu}\|_{\mathrm{TV}}+\lambda\|\bm{Z}\|_{1}\nonumber\\
 &+\langle \bm{C},\bm{\mathcal{F}}^{*}(\bar{\bm{\mu}}-\bm{\mu})+\bar{\bm{Z}}-\bm{Z}\rangle_{F}\geq  \|\bm{\mu}\|_{\mathrm{TV}}+\lambda\|\bm{Z}\|_{1}.
 \end{align}
  Proposition \ref{proposition1} guarantees existence and uniqueness of the solution $(\bm{\mu},\bm{Z})$. The following section shows that the coefficient of $Q(\bm{f})$ can be obtained by solving the dual problem of $\mathrm{P}_{\mathrm{TN}}$.
  \begin{figure*}[t]
  	\centering
  	\mbox{
  		\hspace{-1cm}\subfigure[]{\includegraphics[height=3.6cm]{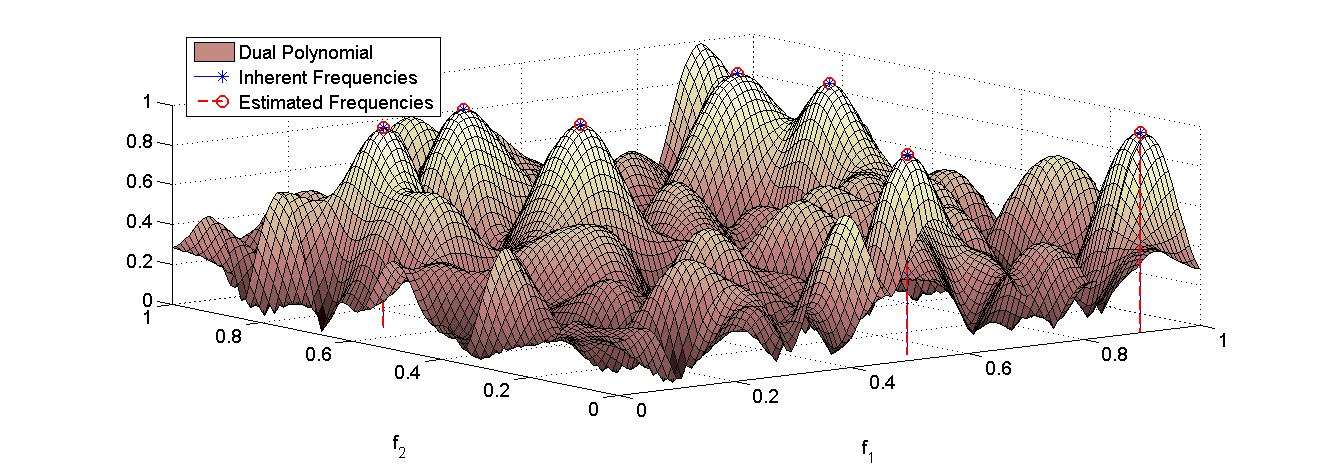}\label{fig.fre}}\hspace{-0.3cm}
  		\subfigure[]{\includegraphics[height=3.6cm]{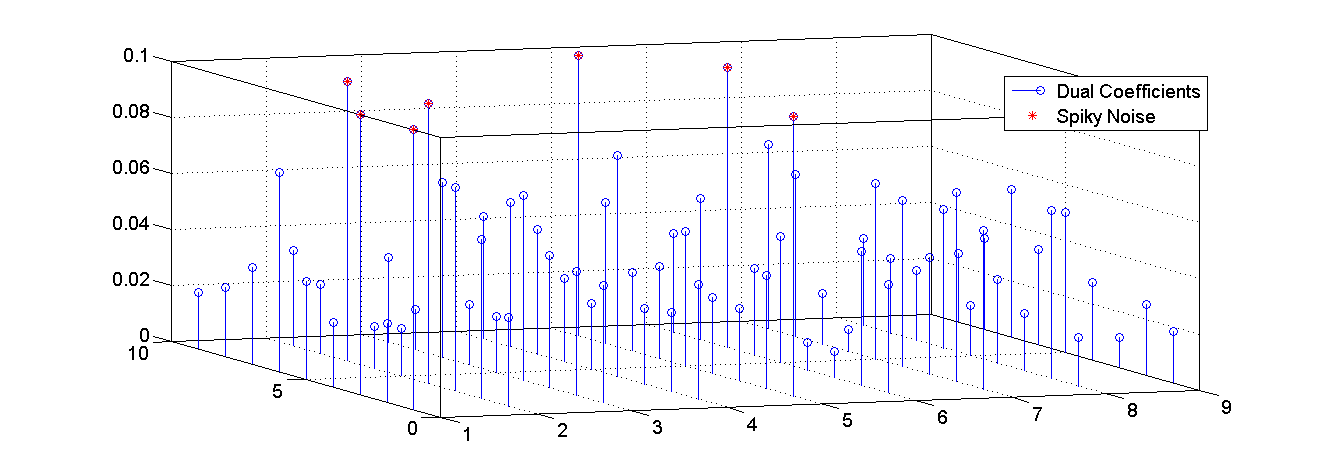}\label{fig.nois}}}
  	\caption{ \ref{fig.fre} shows the magnitude of the dual polynomial corresponding the dual solution of (\ref{sdp}). Also, the support of spectral point sources and estimated sources are represented by blue and red lines, respectively.\ref{fig.nois} demonstrates the magnitude of dual solution of (\ref{sdp}) in blue line and the support of spiky noise in red star.}\label{fig.support}
  \end{figure*}
  \begin{figure*}[t]
  	\centering
  	\mbox{	\hspace{-1cm}
  		\subfigure[]{\includegraphics[width=2.8in]{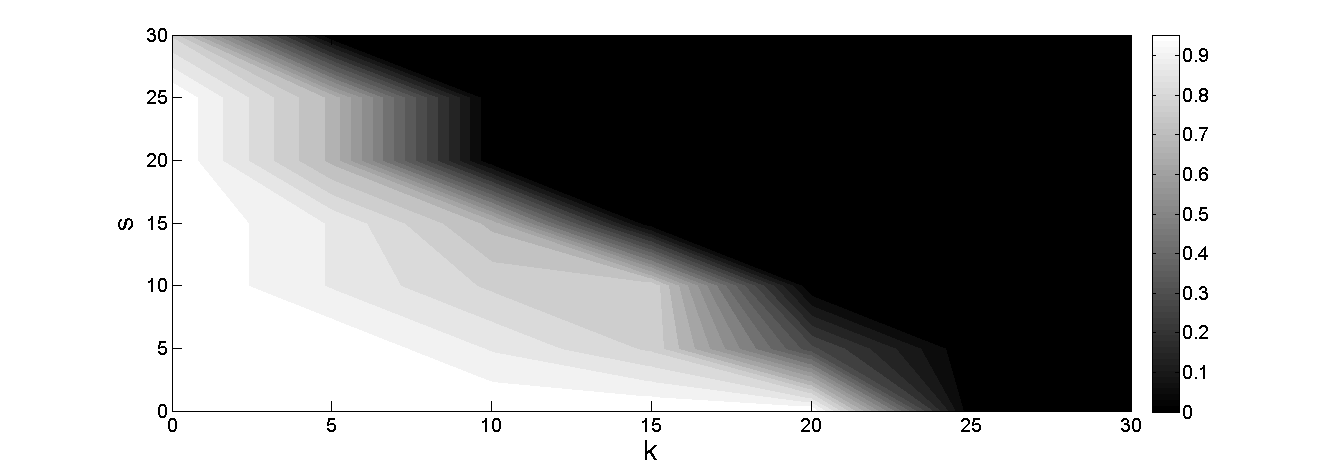}\label{fig.bound4}}\hspace{-1.1cm}
  		\subfigure[]{\includegraphics[width=2.8in]{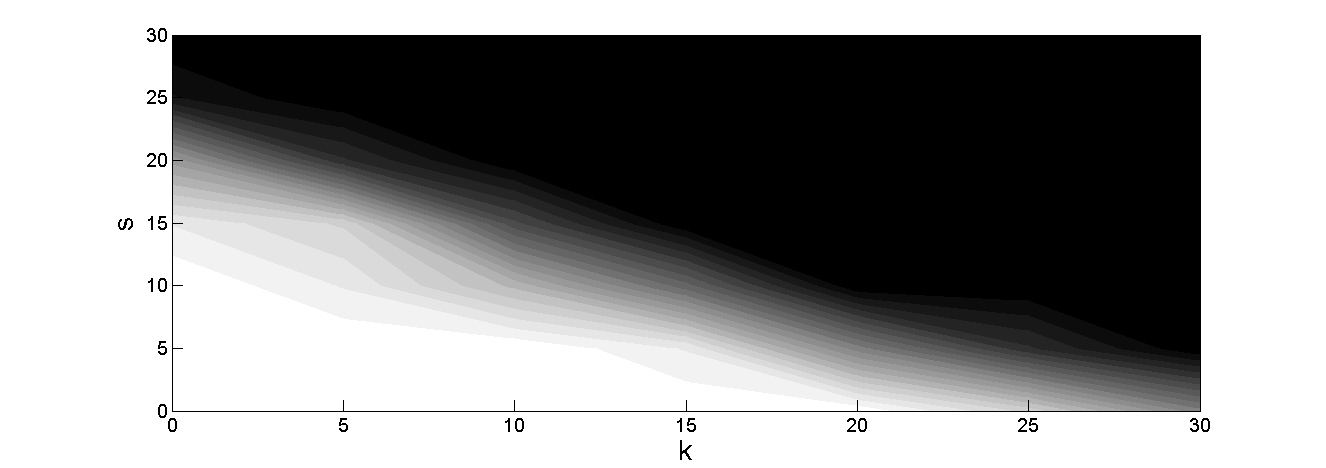}\label{fig.bound4}}\hspace{-0.7cm}
  		\subfigure[]{\includegraphics[width=2.8in]{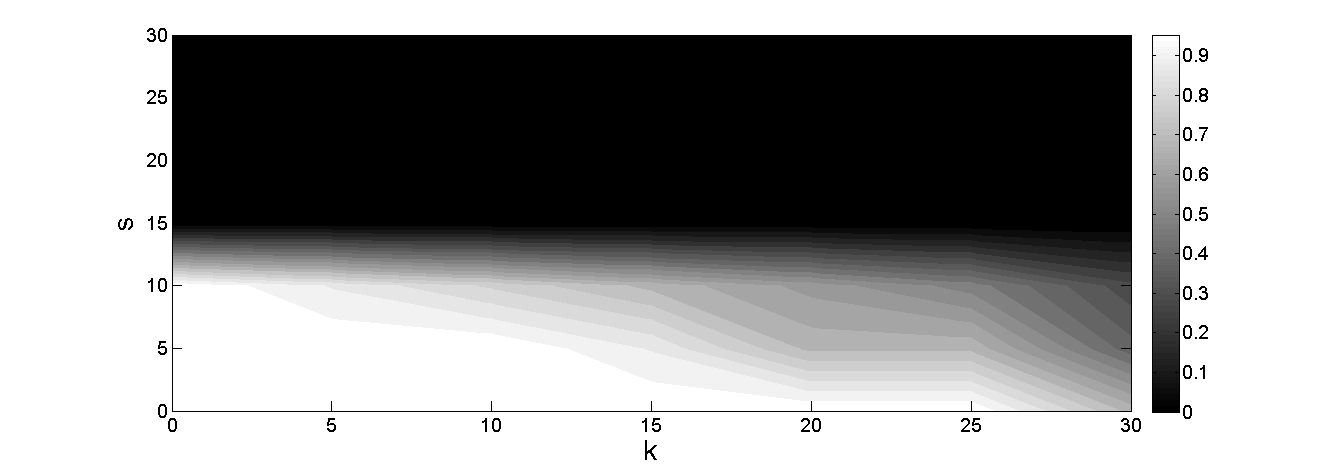}\label{fig.bound5}}}
  	\mbox{
  		\hspace{-1cm}
  		\subfigure[]{\includegraphics[width=2.8in]{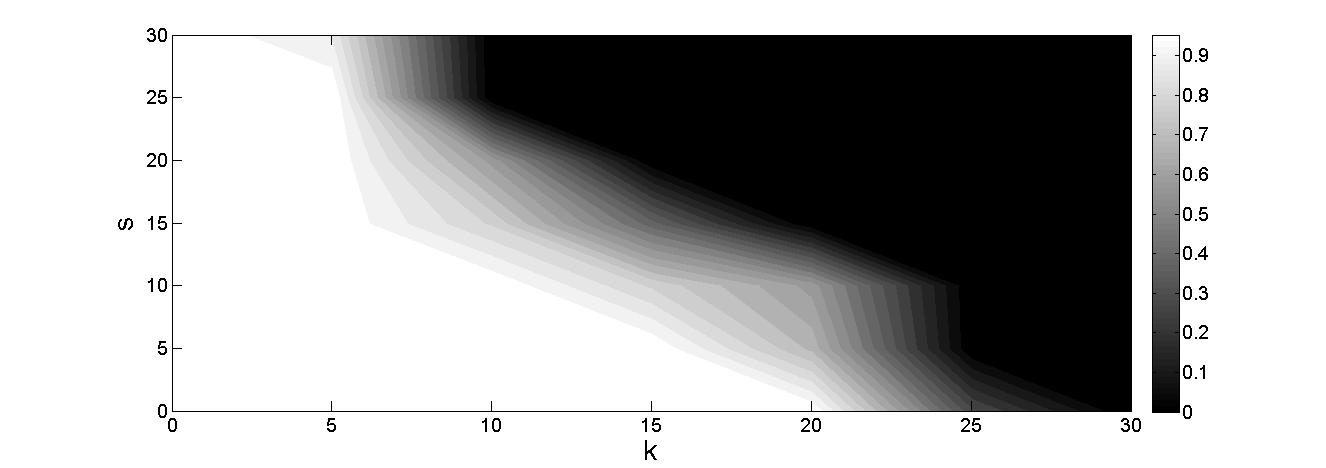}\label{fig.bound4}}\hspace{-1.1cm}
  		\subfigure[]{\includegraphics[width=2.8in]{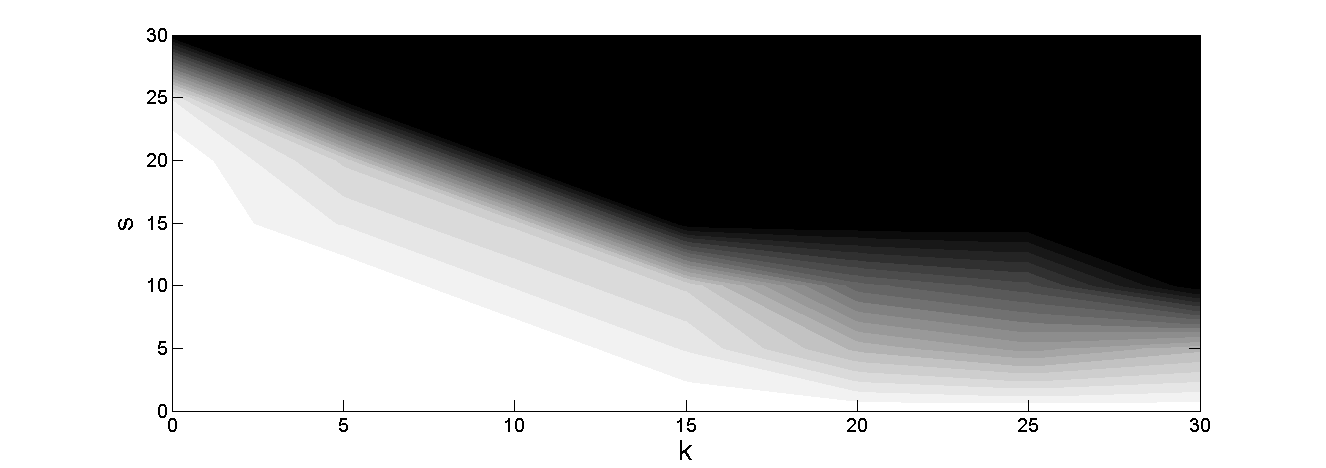}\label{fig.bound4}}\hspace{-0.7cm}
  		\subfigure[]{\includegraphics[width=2.8in]{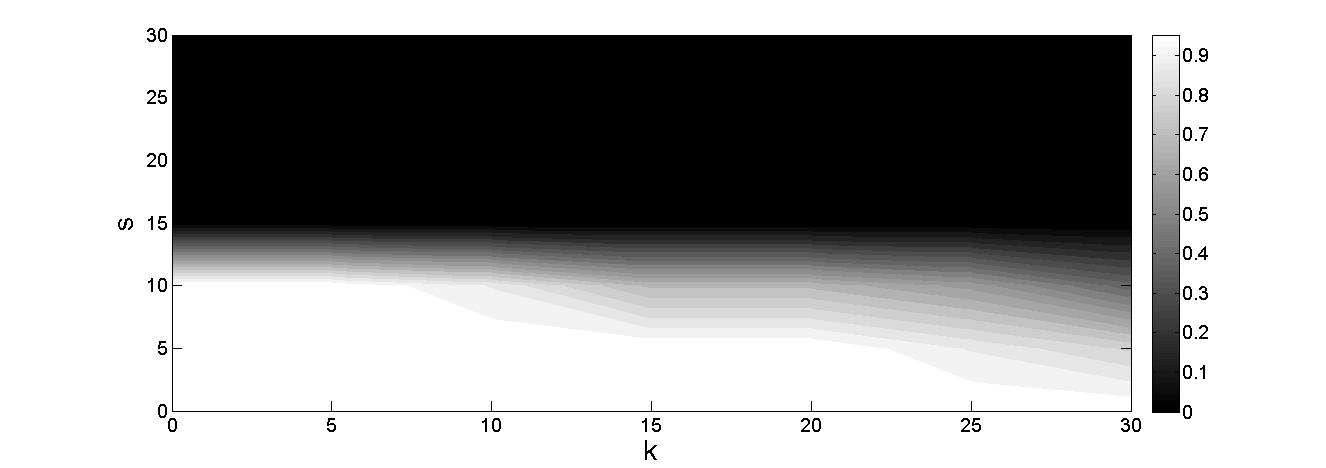}\label{fig.bound5}}}
  	\caption{Grayscale images show the empirical rate of success of (\ref{sdp}) over $10$ trials under the fix minimum separation condition $\frac{3}{n-1}$. First and second rows are corresponding to $n^2=64$ and $n^2=81$, respectively. Also, each column from left to right respectively show results for $\lambda=0.1$, $\lambda=0.125$ and $\lambda=0.2$. }\label{fig.phase}.
  \end{figure*}
   \section{The Dual Problem}\label{imdual}
 The main goal of this section is to suggest a positive semidefinite programming that estimates the location of the frequency sources and the spiky noise. More precisely, we want to convert the infinite dimensional $\mathrm{P}_{\mathrm{TN}}$ into a tractable problem. At the first step, the dual problem of $\mathrm{P}_{\mathrm{TN}}$ can be written using Lagrangian theorem. Then, the results of the PTP theory are applied to convert the explicit constraint of the dual problem into LMI. 
\begin{lem}(Proof in Section \ref{proofdu} )\label{duall}.
	The dual problem of $\mathrm{P}_{\mathrm{TN}}$ is given by
	\begin{align}
	&\max_{\bm{C}\in \mathbb{C}^{n\times n}} \langle\bm{C},\bm{Y}\rangle\nonumber\\ ~~~	&\mathrm{subject~to}~~
 \|\bm{\mathcal{F}}^{*}\bm{C}\|_{\infty}\le 1,~~\|\bm{C}\|_{\infty}\le \lambda,
	\end{align}
\end{lem}
	where $\bm{C}$ is the dual variable. Due to establishment of Slater's condition, there is no gap between the objective value of dual and primal problems \cite{boyd2004convex}. Therefore:  
		\begin{align}\label{kj}
	&\langle\hat{\bm{C}},\bm{Y}\rangle=\langle\hat{\bm{C}},\bm{\mathcal{F}\bm{\mu}+\bm{Z}}\rangle=\|\bm{\mu}\|_{\mathrm{TV}}+\lambda\|\bm{Z}\|_{1},\nonumber\\
	&\langle\bm{\mathcal{F}^{*}}\hat{\bm{C}},\bm{\mu}\rangle+\langle\hat{\bm{C}},\bm{Z}\rangle=\langle\mathrm{sgn}(\bm{\mu}),\bm{\mu}\rangle+\lambda\langle\mathrm{sgn}(\bm{Z}),\bm{Z}\rangle.
	\end{align}
		 Consequently, $|\bm{\mathcal{F}}^{*}\hat{\bm{C}}|=1$ and $|\hat{\bm{C}}_{\bm{k}}|=\lambda$ if $\bm{f}\in T$
and $\bm{k} \in \Omega$, respectively. this provides a strategy to recover the support of spectral sources and spiky noise (see Fig \ref{fig.support}). 
	
The magnitude of the trigonometric polynomial  can be controlled  by LMI using the results of PTP theory \cite{dumitrescu2017positive}. Therefore, one can reformulate the dual problem to positive semidefinite programing as below:
\begin{align}\label{sdp}
\begin{split}
\underset{\bm{C},\bm{Q}_0}{\max} ~~&{\langle \bm{Y},\bm{C}\rangle}\\\nonumber
&\mathrm {subject\,\,to }~ \delta_{\bm{k}}=\mathrm{tr}[\bm{\Theta}_{\bm{k}}\bm{Q}_0],\quad \bm{k}\,\in\ \mathcal{N},
\end{split}\\
&\begin{bmatrix}
\bm{Q}_0 & {\mathrm{vec}(\bm{C})} \\
\\({\mathrm{vec}(\bm{C})})^H & \bm{1}
\end{bmatrix}
\succeq \bm{0},~~\|\bm{C}\|_{\infty}\le \lambda,
\end{align}
where $\bm{Q}_0\in \mathbb{C}^{n^2\times n^2}$ is a Hermitian matrix that $\bm{Q}_{0} \succeq 0$, $\bm{\Theta}_{\bm{k} }=\bm{\Theta}_{k_2 }\otimes\bm{\Theta}_{k_1}$, $\bm{\Theta}_k\in\mathbb{C}^{n\times n} $ is an elementary Toeplitz matrix with ones on it's k-th diagonal and zeros else where, $\delta_{\bm{0}}=1$ and $\delta_{\bm{k}}=\bm{0}$ if $\bm{k}\ne\bm{0}$. In the next section we present numerical experiments to justify (\ref{sdp}).
\section{Experiment}\label{exp}
In this section we present numerical experiments for the observation signal (\ref{noisyoprator}) to investigate the performance of proposed positive semidefint programing (\ref{sdp}). At the first experiment, we  randomly generate $r=7$ frequency sources in $[0,1]^2$ without any separation condition, $s=7$ spiky noises in $\{1,\cdots,n\} \times \{1,\cdots,n\}$ and also the coefficients of each sinusoidal is generated with random magnitude form $\delta_{0.5}+\mathcal{X}^{2}(1)$ and random phase form $\mathcal{U}(0,2\pi)$. We implement (\ref{sdp}) using CVX \cite{grant2008cvx} and leverage the mentioned technique in section \ref{imdual} to recover the support of frequency sources and spiky noise. Fig. \ref{fig.support} demonstrates that local extremums of $|\bm{\mathcal{F}}^{*}\hat{\bm{C}}|$ that achieve one and the locations of $\hat{\bm{C}}$ that achieve $\lambda$ are associated with inherent frequencies of (\ref{superposition}) and locations of spiky noises.\par 
In the second experiment, we investigate the phase transition of proposed approach for different amounts of the regularization parameter $\lambda$, under the fix minimum separation condition $\frac{3}{n-1}$. Indeed, we vary the regularization parameter when varying $k$ and $s$ (Fig. \ref{fig.phase}). As mentioned, $\lambda$ makes a balance between the sparsity of two different components. Small $\lambda$ more strongly penalizes $\mathrm{TV}$ norm of spectral spikes than $\ell_{1}$ norm of spiky noise. This leads to a cost function which is more appropriate to promote the time domain sparsity and vice versa. Fig. \ref{fig.phase} verifies the claim when we change $\lambda$ from small to large values. First and second rows in Fig. \ref{fig.phase} are receptively related to different numbers of measurements $n^2=64$ and $n^2=81$. The estimation is considered successful if the normalized mean squared error $\|\bar{\bm{X}}-\hat{\bm{X}}\|_2/\|\bar{\bm{X}}\|_2\le 10^{-3}$, where $\bar{\bm{X}}$ and $\hat{\bm{X}}$ are associated with $\bm{X}$ and reconstructed data when we omit the corresponding indices of spiky noise, respectively. Grayscale images show the empirical rate of success and each point of those are related to $(k,s,n^2)$. 

\section{Conclusion}\label{Con}
In this work we investigate two-dimensional LSE problem, when a subset of time domain samples is corrupted by spiky noise. In addition, we  proposed a semidefinite programming that achieves the exact solution under mild conditions on the number of spiky noise and the number and separation of spectral sources. One can extend our approach to arbitrary dimensions which is fundamental concern of many applications such as MIMO radar \cite{heckel2017generalized}. It is also worth mentioning that one could consider the model (\ref{noisyproblem}) in compressed sensing regime, namely, in case a random subset of time samples is only observed.
\section{Appendix}
\subsection{Deterministic Certificate}
 Without loss of generality, Assume that $J=\{-m,\cdots,m \} \times \{-m,\cdots\,m\}$ where $m=\frac{n-1}{2}$ and $m=\frac{n}{2}-1$ if $n$ is odd or even, respectively. For dual certificate construction, consider the LSE problem in the noiseless case. \cite{candes2014towards} has shown that the following polynomial can estimate the frequency sources under a mild condition on their separations:
	\begin{align}
\label{dualcertiwithoutoutlier}
\bar{Q}(\bm{f})=\,\sum_{\bm{k}\in J}\bar{C}_{\bm{k}} e^{-j2\pi \bm{f}^T \bm{k}},
\end{align}
such that 
	\begin{align}
&\bar{Q}(\bm{f}_i)=h_{i},  &&\forall \bm{f}_i \in T \label{po11},\\
&|\bar{Q}(\bm{f})| < 1,  &&\forall \bm{f} \notin T.\label{po}
\end{align}
To ensure that $|\bar{Q}(\bm{f})| < 1$ in the off-support, we follow the same approach in \cite{candes2014towards} and \cite{valiulahi2017two} to construct a deterministic dual certificate
\begin{align}
\label{constructed polynimoal}
\bar{Q}(\bm{t})=\sum_{\bm{f}_i \in T}\bar{\alpha}_i\bar{K}(\bm{f}-\bm{f}_i)+\nonumber\bar{\beta}_{1i}\bar{K}^{10}(\bm{f}-\bm{f}_i)\\
+\bar{\beta}_{2i}\bar{K}^{01}(\bm{f}-\bm{f}_i),
\end{align} 
where $\bar{\bm{\alpha}}$, $\bar{\bm{\beta}}_1$ and $\bar{\bm{\beta}}_2$ are interpolation vectors. To meet (\ref{po11}) and (\ref{po}), the following conditions are sufficient
\begin{align}
&\bar{Q}(\bm{f}_i)=h_i,&&\bm{f}_i\in T,\label{signcondition}\\
&\nabla \bar{Q}(\bm{f}_i)=0,&&\bm{f}_i\in T\label{diffcondition}.
\end{align}
\cite{valiulahi2017two} suggested 
\begin{align}
\label{2D kernel}
\bar{K}(\bm{f})=\bar{K}_{\bm{\gamma}}(f_1)\bar{K}_{\bm{\gamma}}(f_2),
\end{align}
for construction, in which
\begin{align}
\label{1Dkernel}
\bar{K}_{\bm{\gamma}}(f)=\prod_{i=1}^{3}K(\gamma_im,f)=\sum_{k=-m}^{m}c_ke^{j2\pi kf},
\end{align}
where $K(\bar{m},f)=\frac{1}{2\bar{m}+1}\sum_{k=-\bar{m}}^{\bar{m}}e^{j2\pi kf}$ is known as Dirichlet kernel, $\gamma_1=0.247$, $\gamma_2=0.339$, $\gamma_3=0.414$, and $\bm{c} \in \mathbb{C}^n$ is the convolution of the Fourier coefficients of $k(\gamma_1m,f)$, $k(\gamma_2m,f)$, and $k(\gamma_3m,f)$. (\ref{signcondition}) and (\ref{diffcondition}) can be reformulated as a matrix equation 
\begin{align}
\underset{\bar{\bm{E}}}{\underbrace{\begin{bmatrix}
	\bar{\bm{E}}_{00} & \kappa\bar{\bm{E}}_{10} & \kappa\bar{\bm{E}}_{01}\\
	-\kappa\bar{\bm{E}}_{10} & -\kappa^2\bar{\bm{E}}_{20} & -\kappa^2\bar{\bm{E}}_{11}\\
	-\kappa\bar{\bm{E}}_{01} & -\kappa^2\bar{\bm{E}}_{11} & -\kappa^2\bar{\bm{E}}_{02}\\
	\end{bmatrix}}}
\begin{bmatrix}
\bar{\bm{\alpha}}\\
\kappa^{-1}\bar{\bm{\beta}}_1\\ 
\kappa^{-1}\bar{\bm{\beta}}_2
\end{bmatrix}
=
\begin{bmatrix}
\bm{h}\\
\bm{0}\\ 
\bm{0}
\end{bmatrix},
\end{align}
where
$(\bar{\bm{E}}_{i_1i_2})_{\ell,j}=\bar{K}^{(i_1i_2)}(\bm{f}_\ell-\bm{f}_j)$
and $\kappa:=\frac{1}{\sqrt{|K^{''}|(0)}}$. We borrow two bounds on $\|\bm{c}\|_\infty$ and $\kappa$ form \cite{fernandez2016super} which are useful in advancing our proof
\begin{align}
&\|\bm{c}\|_{\infty}\le \frac{1.3}{m},\label{boundonc}\\
&\frac{0.467}{m}\le \kappa\le \frac{0.468}{m},~~~ \text{for}~ m\geq 2\times 10^3.		\label{boundonkappa}
\end{align}
\subsection{Random Certificate}
Spiky noise randomly corrupts a subset of time samples, it is similar to random sampling in compressed sensing literature \cite{tang2013compressed}. \cite{fernandez2016demixing} uses the same technique to incorporate the randomness of spiky noise into dual certificate construction. We follow this approach to construct a valid certificate for two-dimensional case when a  subset of time samples dose not follow the exponential structure. First, $Q(\bm{f})$ is divided into two terms 
\begin{align}
\label{decompose}
Q(\bm{f}) :=Q_{\text{aux}}(\bm{f})+R(\bm{f}),
\end{align}
where 
\begin{align}
\label{Q_aux}
&Q_{\text{aux}}(\bm{f}):=\sum_{\bm{k}\in\Omega^c}C_{\bm{k}}e^{-j2\pi\bm{f}^T\bm{k}},\\
\label{R}
&R(\bm{f}):=\frac{1}{\sqrt{n^2}}\sum_{\bm{k}_{l}\in\Omega}{r}_{l}e^{-j2\pi\bm{f}^T\bm{k}_{l}}.
\end{align}
 The coefficients of the first term are restricted to $\Omega^{c}$ and the coefficients of the second term are in $\Omega$. From the definitions, it is obvious that (\ref{con3}) is satisfied and $R(\bm{f})$ has no degree of freedom. So, we construct $Q_{\text{aux}}(\bm{f})$ to guarantee other conditions in proposition \ref{proposition1}. The inequality $|Q(\bm{f})|\le 1, ~\forall \bm{f}_i \in T$ is satisfied by setting to zero the partial derivatives at $T$. So
\begin{align}
\label{constructedployy}
&Q_{\text{aux}}(\bm{f}_i)=h_{i}-R(\bm{f}_i), &&\forall \bm{f}_i\in T,\\
&\nabla Q_{\text{aux}}(\bm{f}_i)=-\nabla R(\bm{f}_i),&&\forall\bm{f}_i\in T.
\label{constructedployy1}
\end{align}
Define a restricted version of $\bar{K}$ on $\Omega^{c}$
\begin{align}
\label{sampled2dkernel}
&K(\bm{f}):=\sum_{\bm{k}\in\Omega^c}c_{k_1}c_{k_2}e^{j2\pi\bm{f}^T\bm{k}}=\sum_{\bm{k}\in J}\delta_{\Omega^c}(\bm{k})c_{k_1}c_{k_2}e^{j2\pi\bm{f}^T\bm{k}},
\end{align}
where $\delta_{\Omega^c}(\bm{k})=1$ if $\bm{k}=\Omega^{c}$, $\delta_{\Omega^c}(\bm{k})=0$ otherwise. Under the noise condition of Theorem \ref{thm.main}, these are independent Bernoulli random variables with parameter $\frac{n^2-s}{n^2}$. So the expectation of $K(\bm{f})$ can be written as:
\begin{align}
\label{themeanofK}
&\mathbb{E}(K(\bm{f}))=\frac{n^2-s}{n^2}\sum_{k\in J}c_{k_1}c_{k_2}e^{j2\pi\bm{f}^T\bm{k}}=\frac{n^2-s}{n^2}\bar{K}(\bm{f}).
\end{align}
The mean of partial derivatives of $K(\bm{f})$ can be obtained by the same technique. we construct $Q_{\text{aux}}$ by a linear combination of $K(\bm{f})$ and its partial derivatives:
\begin{align}
\label{constructed random polynimoal}
Q_{\text{aux}}(\bm{t})=\sum_{\bm{f}_i \in T}\alpha_iK(\bm{f}-\bm{f}_i)+\nonumber\beta_{1i}K^{10}(\bm{f}-\bm{f}_i)\\
+\beta_{2i}K^{01}(\bm{f}-\bm{f}_i),
\end{align} 
where $\bm{\alpha},~\bm{\beta}_{1}~\text{and} ~\bm{\beta}_{2} \in \mathbb{C}^{|T|}$ are interpolation coefficient vectors. (\ref{constructedployy}) and (\ref{constructedployy1}) can be recast in matrix form as follows 
\begin{align} 
\label{linearsystem} 
\underset{\hspace{-1.5cm}\bm{E}}{\underbrace{\begin{bmatrix}
	\bm{E}_{00} & \kappa\bm{E}_{10} & \kappa\bm{E}_{01}\\
	-\kappa\bm{E}_{10} & -\kappa^2\bm{E}_{20} & -\kappa^2\bm{E}_{11}\\
	-\kappa\bm{E}_{01} & -\kappa^2\bm{E}_{11} & -\kappa^2\bm{E}_{02}\\
	\end{bmatrix}}
	\begin{bmatrix}
	\bm{\alpha}\\
	\kappa^{-1}\bm{\beta}_1\\ 
	\kappa^{-1}\bm{\beta}_2
	\end{bmatrix}}
=
\begin{bmatrix}
\bm{h}\\
\bm{0}\\ 
\bm{0}
\end{bmatrix}
-\frac{1}{\sqrt{n^2}}B_{\Omega}\bm{r}
\end{align}
where $(\bm{E}_{i_1i_2})_{\ell,j}=K^{(i_1i_2)}(\bm{f}_\ell-\bm{f}_j)$ and $\frac{1}{\sqrt{n^2}}B_{\Omega}\bm{r}$ can be written versus the components of $R(\bm{f})$ and their partial derivatives
\begin{align}
\label{BRomega}
\frac{1}{\sqrt{n^2}}B_{\Omega}\bm{r}=\big[R(\bm{f}_1)...R(\bm{f}_k),~R^{10}(\bm{f}_1)...R^{10}(\bm{f}_k),\nonumber\\R^{01}(\bm{f}_1)...R^{01}(\bm{f}_k)\big]^T.
\end{align}
Define
\begin{align}
\label{bk}
&B_\Omega:=\big[\bm{b}(\bm{k}_{i_1}),\cdots,\bm{b}(\bm{k}_{i_s})\big],~~~~\Omega=\{i_1,\cdots,i_s\},\nonumber\\
&\bm{b}(\bm{k})=
\begin{bmatrix}
1\\-j2\pi\kappa k_1\\-j2\pi\kappa k_2
\end{bmatrix}
\otimes
\begin{bmatrix}
e^{-j2\pi\bm{f}_{1}^T\bm{k}}\\
\cdot\\
e^{-j2\pi\bm{f}_{r}^T\bm{k}}
\end{bmatrix},
\end{align}
where  $\bm{k}_{i_1},\cdots,\bm{k}_{i_s}$ are associated with $\bm{k}\in \Omega$. Interpolation vectors can be computed by solving the linear system  (\ref{linearsystem}), so 
\begin{align}
\label{Qfinv}
Q(\bm{f})=w^{00}(\bm{f})^TE^{-1}\bigg(\begin{bmatrix}
\bm{h}\\
\bm{0}\\ 
\bm{0}
\end{bmatrix}-\frac{1}{\sqrt{n}}B_{\Omega}\bm{r}\bigg)+R(\bm{f}),
\end{align}
where $w^{i_1i_2}(\bm{f})$ for $i_1,i_2~\in\{0,1,2\}$ is defined  
\begin{align}
\label{wf}
&w^{i_1i_2}(\bm{f}):=\kappa^{i_1+i_2}\bigg[K^{i_1i_2}(\bm{f}-\bm{f}_1),\cdots,K^{i_1i_2}(\bm{f}-\bm{f}_k),\nonumber\\&\kappa K^{i_1+1,i_2}(\bm{f}-\bm{f}_1),\cdots,\kappa K^{i_1+1,i_2}(\bm{f}-\bm{f}_k)\nonumber\\&\kappa K^{i_1,i_2+1}(\bm{f}-\bm{f}_1),\cdots,\kappa K^{i_1,i_2+1}(\bm{f}-\bm{f}_k)\bigg]^T.
\end{align}
The following  Lemma establishes an upper bound on the $\ell_{2}$ norm of $\bm{b}(\bm{k})$.
\begin{lem}\label{boundonb}
If $m \geq 2\times10^3$, then
\begin{align}
\|\bm{b}(\bm{k})\|^2_2\le21~r, ~~\text{for}~~\bm{k}\in J.
\end{align}
\end{lem}
Proof. 
\begin{align}
\|\bm{b}(\bm{k})\|^2_2&\le r\Big(1+\max_{|k_1|\le m}(2\pi k_1 \kappa)^2+\max_{|k_2|\le m}(2\pi k_2 \kappa)^2\Big)\nonumber\\&\le21r.
\end{align}
The following lemma establishes an upper bound on the operator norm of $\bm{B}_{\Omega}$ with certain probability.
\begin{lem}\label{boundbomega}
	(proof in section \ref{proofoflemmaboundomega}). If the conditions of Theorem \ref{thm.main} hold, the event 
	\begin{align}
\varepsilon_{B}:=\bigg\{\|\bm{B}_\Omega\|>C_B(\log\frac{n^2}{\epsilon})^{-1/2}\sqrt{n^2}\bigg\},
	\end{align}
	happens with probability $\frac{\epsilon}{5}$ in which numerical constant $C_B$ is defined in (\ref{CBdefine}).
	\end{lem}
The following lemma sates that  $w^{i_1i_2}(\bm{f})$ is concentrated around the scales version of \begin{align}
\label{wfbar}
&\bar{w}^{i_1i_2}(\bm{f}):=\kappa^{i_1+i_2}\bigg[\bar{K}^{i_1i_2}(\bm{f}-\bm{f}_1)~~...~~\bar{K}^{i_1i_2}(\bm{f}-\bm{f}_k)\nonumber\\&\kappa \bar{K}^{i_1+1,i_2}(\bm{f}-\bm{f}_1)~~...~~\kappa \bar{K}^{i_1+1,i_2}(\bm{f}-\bm{f}_k)\nonumber\\&\kappa \bar{K}^{i_1,i_2+1}(\bm{f}-\bm{f}_1)~~...~~\kappa \bar{K}^{i_1,i_2+1}(\bm{f}-\bm{f}_k)\bigg]^T,
\end{align}
on a fine grid with high probability.
\begin{lem}\label{event}(Proof in section \ref{proofboundw}). Let $\mathcal{G} \subset [0,1]^2$ be a two-dimensional equispaced $800n^4$ grid that discretizes $[0,1]^2$. If the conditions of Theorem \ref{thm.main} hold, then the event
\begin{align}
\varepsilon_{v}:=\bigg\{\bigg\|w^{i_1i_2}(\bm{f})-\frac{n^2-s}{n^2}\bar{w}^{i_1i_2}(\bm{f})\bigg\|_2>C_v(\log\frac{n^2}{\epsilon})^{-1/2}\bigg\},
\end{align}
happens with probability $\epsilon/5$ for all $\bm{f} \in \mathcal{G}$, $i_1,i_2 \in \{0,1,2,3\}$ and  numerical constant $C_v$ in (\ref{defccbb}).
\end{lem}
\subsection{Proof of Proposition \ref{proposition1} }
In the first step, we seek to determine the uniqueness of the solution of linear system (\ref{linearsystem}). 
The following lemma shows that $\bm{E}$
is concentrated around $\bar{\bm{E}}$
with high probability. Consequently, $\bm{E}$ is invertible and one can bound the operator norm of its inverse.
\begin{lem}\label{inverse result}(Proof in section \ref{proofboundebar}). If the conditions of Theorem \ref{thm.main} hold, then the event 
			\begin{align}
	\varepsilon_{E}:=\Bigg\{ \Big\|\bm{E}-&\frac{n^2-s}{n^2}\bar{\bm{E}}\Big\|\nonumber\\
	&
	\geq\frac{n^2-s}{4n^2}\min\bigg\{1,\frac{C_D}{4}\bigg\}\Big(\log\frac{n^2}{\epsilon}\Big)^{\frac{-1}{2}}\Bigg\},
	\end{align}
	happens with probability $\epsilon/5$. Also, under the event $\varepsilon_{E}^c$, $\bm{E}$ is invertible and 
		\begin{align}
&\big\|\bm{E}^{-1}\big\| \le 8,\nonumber\\
&\Big\|\bm{E}^{-1}-\frac{n^2}{n^2-s}\bar{\bm{E}}^{-1}\Big\|\le C_D\big(\log\frac{n^2}{\epsilon}\big)^{\frac{-1}{2}},
	\end{align}
	where $C_D$ is the numerical constant which is defined by (\ref{definecd}). \end{lem}
Indeed, this Lemma states that, under the event  $\varepsilon^c_{D}$, the linear system (\ref{linearsystem}) has an stable solution. So $Q(\bm{f})$ is well defined and (\ref{con1}) holds. In order to meet (\ref{con2}), it is sufficient to show that $Q(\bm{f})$ is concentrated around $\bar{Q}(\bm{f})$ on a fine gird. After that,  using Bernstein's inequality, we demonstrate that this holds on the whole $[0,1]^2$. Finally we borrow some bound on $\bar{Q}(\bm{f})$ and its partial derivative from \cite{valiulahi2017two} to complete the proof.
\begin{lem}(Proof in section \ref{proofprop2}).\label{prop2}
	If the conditions of theorem \ref{thm.main} hold, then
	\begin{align}
		|Q(\bm{f})|<1,~~ \text{for}~~\bm{f} \notin T,
	\end{align}
 with probability $1-\epsilon/5$ under the event $\varepsilon_{B}^c \cap \varepsilon_{E}^c \cap \varepsilon_{v}^c $.
\end{lem}
The last to show is (\ref{con4}). The following lemma states that under the event
	 $\varepsilon_{B}^c \cap \varepsilon_{E}^c \cap \varepsilon_{v}^c $, one can control the magnitude of dual polynomial's coefficients with high probability. 
			\begin{lem}(Proof in section \ref{ProoflemmaC}\label{prop3}). If the conditions of Theorem \ref{thm.main} hold, then
		\begin{align}
		|C_{\bm{k}}|<\frac{1}{\sqrt{n^2}},~~~\text{for}~~ \bm{k} \in \Omega^{c},
		\end{align}
		 under the event $\varepsilon^c_{B}\cap\varepsilon^c_{D}\cap\varepsilon^c_{v}$.
	\end{lem}
Finally, we use the same technique in \cite{fernandez2016demixing} to complete the proof. Consider $\varepsilon_{Q}$ and $\varepsilon_{q}$  as the events, such that (\ref{con2}) and (\ref{con4}) hold, respectively. By De Morgan's laws and union bound we have 
\begin{align}
\hspace{-0.1cm}\mathbb{P}((\varepsilon_{Q} \cap \varepsilon_{q})^{c})&=\mathbb{P}(\varepsilon_{Q}^{c} \cup \varepsilon_{q}^{c})\nonumber\\
&\le \mathbb{P}(\varepsilon_{Q}^{c} \cup \varepsilon_{q}^{c}|\varepsilon^c_{B}\cap\varepsilon^c_{D}\cap\varepsilon^c_{v})+\mathbb{P}(\varepsilon_{B}\cap\varepsilon_{E}\cap\varepsilon_{v})\nonumber\\
&\le \mathbb{P}(\varepsilon_{Q}^{c}|\varepsilon^c_{B}\cap\varepsilon^c_{D}\cap\varepsilon^c_{v})+\mathbb{P}(\varepsilon_{q}^{c}|\varepsilon^c_{B}\cap\varepsilon^c_{D}\cap\varepsilon^c_{v})\nonumber\\
&+\mathbb{P}(\varepsilon_{B})+\mathbb{P}(\varepsilon_{E})+\mathbb{P}(\varepsilon_{v}) \nonumber\\ & \le \epsilon, 
\end{align}
which holds by the fact that for any pair of events $\varepsilon_{A}$ and $\varepsilon_{B}$ we have $\mathbb{P}(\varepsilon_{A}) \le \mathbb{E}(\varepsilon_{A}|\varepsilon_{B}^{c})+\mathbb{E}(\varepsilon_{B})$. On the other hand, via Lemmas \ref{prop3}, \ref{prop3}, \ref{inverse result}, \ref{event} and \ref{boundbomega}, it is shown that the construction is valid with probability at least $1-\epsilon$.
\section{Proof of Lemma \ref{boundbomega}}\label{proofoflemmaboundomega}
In order to obtain an upper bound on the operator norm of $B_{\Omega}$, under the assumption of theorem \ref{thm.main}, we show that 
\begin{align}
\bm{H}:=B_{\Omega}B^{*}_{\Omega}=\sum_{\bm{k}\in\Omega}\bm{b}(\bm{k})\bm{b}^{*}(\bm{k}),
\end{align}
is concentrated around 
\begin{align}
\frac{s}{n^2}\bar{\bm{H}}=\frac{s}{n^2}\sum_{\bm{k}\in J}\bm{b}(\bm{k})\bm{b}^{*}(\bm{k}).
\end{align}
 Using the following lemma, we can compute an upper bound on the operator norm of $\bar{\bm{H}}$.
\begin{lem}(Proof in section \ref{ProofHbar}).\label{boundonH}
	If the conditions of Theorem \ref{thm.main} hold, then
	\begin{align}
	\|\bar{\bm{H}}\|\le 223707~n^2\log^2k.
	\end{align}
\end{lem}
Regarding the fact that $s\le C_sn^2\big(\log^2 k~\log \frac{n^2}{\epsilon}\big)^{-1}$ in Theorem \ref{thm.main} we have  
\begin{align}\label{boundonHbar}
	\|\frac{s}{n^2}\bar{\bm{H}}\|\le \frac{C^2_{B}}{2}n^2\big(\log\frac{n^2}{\epsilon}\big)^{-1},
	\end{align}
	if we set $C_{B}$ small enough. One can demonstrate that $\bm{H}$ concentrates around the scaled version of $\bar{\bm{H}}$ using matrix Bernstein inequality.
	\begin{lem}(Proof in section \ref{proflem7}). \label{boundonHandHbar}
		If the conditions of Theorem \ref{thm.main} hold, then
		\begin{align}
		\|\bm{H}-\frac{s}{n^2}\hat{\bm{H}}\|\le\frac{C^2_B}{2}n^2\big(\log\frac{n^2}{\epsilon}\big)^{-1}
		\end{align}
		with probability at least $1-\epsilon/5$.
	\end{lem}
	Consequently, one can bound the operator norm of $B_{\Omega}$ by triangle inequality
		\begin{align}
	\|B_{\Omega}\|&\le\sqrt{\|\bm{H}\|}\le\sqrt{\|\frac{s}{n^2}\bar{\bm{H}}\|+\|\bm{H}-\frac{s}{n^2}\bar{\bm{H}}\|}\nonumber\\&\le C_B\sqrt{n^2}\big(\log \frac{n^2}{\epsilon}\big)^{-1/2},
	\end{align}
	which happens with probability at least $1-\epsilon/5$.
	\section{Proof of Lemma \ref{boundonH}}\label{ProofHbar}
	 Consider two-dimensional Dirichlet kernel  
\begin{align}\label{2dkernel}
\bar{K}(\bm{f}):=\frac{1}{n^2}\sum_{\bm{k}\in J}e^{j2\pi\bm{f}^T\bm{k}}=\bar{K}(f_1)\bar{K}(f_2),
\end{align}
where $\bar{K}_{m}(f):=\frac{1}{n}\sum_{k\in \{-m,...,m\}}e^{j2\pi fk}$ is known as Dirichlet kernel.
	 $\bar{\bm{H}}$ can be recast in matrix form using $\bar{K}(\bm{f})$ as 
	\begin{align}
		 \bar{\bm{H}} :=n^2 
	\begin{bmatrix}
	\bar{\bm{H}}_{00} & \kappa\bar{\bm{H}}_{10} & \kappa\bar{\bm{H}}_{01}\\
	-\kappa\bar{\bm{H}}_{10} & -\kappa^2\bar{\bm{H}}_{20} & -\kappa^2\bar{\bm{H}}_{11}\\
	-\kappa\bar{\bm{H}}_{01} & -\kappa^2\bar{\bm{H}}_{11} & -\kappa^2\bar{\bm{H}}_{02}\\
	\end{bmatrix}
		\end{align}
	where $
	\label{Hbar}
	(\bar{\bm{H}}_{i_1i_2})_{\ell,j}=\bar{K}^{i_1i_2}(\bm{f}_\ell-\bm{f}_j)
$. A uniform bound on the magnitude of the Dirichlet kernel is obtained by Bernstein's polynomial inequality in \cite{fernandez2016super}
		\begin{align}
	\label{boundon1-d}
|\bar{K}^{\ell}_{m}(f)| \le (2m)^{\ell}.
	\end{align}
	So we can uniformly bound the magnitude of $\bar{K}(\bm{f})$ and its partial derivatives using this bound and (\ref{2dkernel})
			\begin{align}
	\label{boundon2-d}
	|\bar{K}^{i_1i_2}(\bm{f})| \le (2m)^{i_1+i_2}.
	\end{align}
	There is another bound on the magnitude of Dirichlet kernel and its derivatives which holds for $m\geq 10^3$ and $f\geq 80/m$ 
				\begin{align}
	\label{decreasingboundon1-d}
|\bar{K}^{\ell}_{m}(f)| \le \frac{1.1~2^{\ell-2}\pi^\ell m^{\ell -1}}{f}
	\end{align}
 (See section C.4 for proof). Similar to (\ref{2dkernel}), we have 
 	\begin{align}
 \label{decboundon2-d}
 |\bar{K}^{i_1i_2}(\bm{f})| \le \frac{(1.1)^2~2^{i_1+i_2-4}\pi^{i_1+i_2}m^{i_1+i_2-2}}{f_1f_2},
 \end{align}
 for the domain in which $\min({|f_1|,|f_2|)}\geq 80/m$. 
 
 These bounds can be used for bounding sum of magnitudes of $\bar{K}(\bm{f})$ and its partial derivatives. Assume that $\bm{f}_i \in T$ is fixed. There are at most $95^{2}$ other frequency sources that are in square $\|\bm{f}-\bm{f}_i\|_{\infty}\le 80/m$ with respected to the minimum $\frac{3.36}{m}$. Now we are able to bound those terms as below
  	\begin{align}
 \label{boundonsum} 
 &\sup_{\bm{f}_i}\sum_{j=1}^{k}\kappa^{i_1+i_2}|\bar{K}^{i_1i_2}(\bm{f}_i-\bm{f}_j)|\le95^2\kappa^{i_1+i_2}\sup_{\bm{f}}|\bar{K}^{i_1i_2}(\bm{f})|\nonumber\\
 &+\kappa^{i_1+i_2}\sum_{j=1}^{k}\sup_{\min({|f_1|,|f_2|)}\geq j\Delta_{\min}}\hspace{-1cm}|\bar{K}^{i_1i_2}(\bm{f})|\le 95^2\nonumber\\&
 +\frac{(1.1)^2}{m^{i_1+i_2}}\sum_{i=1}^{k}\sum_{j=1}^{k}\frac{\pi^{i_1+i_2}m^{i_1+i_2-2}}{16~ij\Delta_{\min}^2}\le 74569\log^2k,
 \end{align}
 for $k\geq2$ and $i_1, i_2 \in \{0,1,2,3\}$. The last inequality is obtained by $\Delta_{\min}=1.68/m$, $\sum_{i=1}^{k}\frac{1}{i}\le1+\log k\le 2\log k$, (\ref{boundonkappa}), and the fact that $9025+108\log^2k\le 74569\log^2k$.
 It is possible to bound the eigenvalue of $\bar{\bm{H}}$ using Gershogrin's disk theorem which leads to a bound on the operator norm of $\bar{\bm{H}}$. 
 	\begin{align}
 \label{boundonopratorH}
 &\|\bar{\bm{H}}\| \le n^2 \max_{i}\bigg\{\sum_{j=1}^{k}|\bar{K}(\bm{f}_i-\bm{f}_j)|+2\sum_{j=1}^{k}\kappa|\bar{K}^{10}(\bm{f}_i-\bm{f}_j)|,\nonumber\\ &\sum_{j=1}^{k}\kappa\ |\bar{K}^{10}(\bm{f}_i-\bm{f}_j)|+\sum_{j=1}^{k}\kappa^2|\bar{K}^{20}(\bm{f}_i-\bm{f}_j)|\nonumber\\&+\sum_{j=1}^{k}\kappa^2|\bar{K}^{11}(\bm{f}_i-\bm{f}_j)| \bigg\}\le 223707~n^2\log^2k.
 \end{align}
 This concludes the proof.
 \section{Proof of Lemma \ref{boundonHandHbar}}\label{proflem7}
 \begin{lem} \label{matrixbernsitann}(Bernstein's matrix inequality \cite{tropp2012user}). Let $\bm{X}({1}),\cdots,\bm{X}({L})$ be independent zero-mean self-adjoint random matrices of dimension $d\times d$. If $\|\bm{X}(k)\|\le B ~\forall k$, we have
 	 	\begin{align}
 	\mathbb{P}\bigg\{\big\|\sum_{k=1}^{L}\mathbb{E}\big[\bm{X}(k)\big]\big\|\geq t\bigg\}\le 2d\exp\big(\frac{-t^2/2}{\sigma^2+Bt/3}\big),
 	\end{align}
 	 for any $t \geq 0$ and  $\sigma^2:=\big\|\sum_{k=1}^{L}\mathbb{E}\big[\bm{X}(k)\bm{X}^{T}(k)\big]\big\|$.
 	 \end{lem}
  Consider a finite sequence of independent adjoint zero-mean random matrices of the form 
  	\begin{align}
 \label{zero-mean}
 \bm{X}({\bm{k}}):=\big(\delta_{\Omega}(\bm{k})-\frac{s}{n^2})\bm{b}(\bm{k})\bm{b}^{*}(\bm{k}),~~~~ \bm{k}\in J.
 \end{align}
 The aim is to show that $\bm{H}$ concentrates around $\frac{s}{n^2}\bar{\bm{H}}$ with high probability.
 We can write
   	\begin{align}
\bm{H}-\frac{s}{n^2}\bar{\bm{H}}=\sum_{\bm{k}\in J}\bm{X}(\bm{k}).
 \end{align}
 One can bound the operator norm of $\bm{X}$ by Lemma \ref{boundonb}
    	\begin{align}
    	\label{boundonopratorX}
    	\|\bm{X}(\bm{k})\|\le \max_{\bm{k}\in J} \|\bm{b}(\bm{k})\|^2_2 \le B:=21r.
 \end{align}
 Also,
 	\begin{align}
 \label{boundonsigma}
 \sigma^2&:=\Big\|\sum_{\bm{k}\in J}\|\bm{b}(\bm{k})\|_2^2~\bm{b}(\bm{k})\bm{b}^*(\bm{k})\mathbb{E}\big[\big(\delta_{\Omega}(\bm{k})-\frac{s}{n^2}\big)^2\big]\Big\|\nonumber\\
 &\le \frac{21rs}{n^2}\|\bar{\bm{H}}\|\le \frac{21C^2_Bn^2r}{2}\bigg(\log\frac{n^2}{\epsilon}\bigg)^{-1},
 \end{align}
 where the last inequality comes from (\ref{boundonHbar}), the second from Lemma \ref{boundonb} and the first stems from the variance of  Bernoulli model with parameter $\frac{s}{n^2}$. Assume $t:=\frac{C^2_Bn^2}{2}\Big(\log\frac{n^2}{\epsilon}\Big)^{-1}$ in Lemma \ref{matrixbernsitann} for simplicity, so $\sigma^2:=21rt$. We can write 
 	\begin{align}
 \label{bernes}
 \mathbb{P}\Big[\|\mathbf{H}-\frac{s}{n^2}\bar{\bm{H}}\|\geq t\Big]\le 3r \exp \Big(\frac{-3t}{168r}\Big).
 \end{align}
 The lower bound of the probability is equal to $\epsilon/5$ under the condition
 	\begin{align}
 \label{boundonk}
 r\le \frac{3C^2_Bn^2}{336}\Big(\log\frac{n^2}{\epsilon}\log \frac{15r}{\epsilon}\Big)^{-1}. 
 \end{align}
 This criterion is satisfied by setting $C_r$ in Theorem \ref{thm.main} small enough. 
 \section{Proof Lemma \ref{event}}\label{Proofboundonw}
 \begin{lem} \label{vecbern}(Vector Bernstein's inequality \cite{talagrand1995concentration}).
 Let $\bm{u}(1),\cdots,\bm{u}(L)$ be independent zero-mean random vectors of dimension $d$. If $\|\bm{u}(k)\|_{2}\le B ~\forall k$, we have
 \begin{align}
 \mathbb{P}\bigg\{\sum_{k=1}^{L}\|\bm{u}(k)\|_{2}\geq t\bigg\}\le \exp\big(-\frac{t^2}{8\sigma^2}+\frac{1}{4}\big),
 \end{align}
 for any $0 \le t \le \sigma^2$ where $\sum_{k=1}^{L}\mathbb{E}\big[\|\bm{u}(k)\|^2_{2}\big]\le \sigma^2$.
\end{lem}
 Let us write $\bar{w}^{i_1i_2}(\bm{f})$ and $w^{i_1i_2}(\bm{f})$ in the term of $\bm{b}$ using the definition of $\bar{K}(\bm{f})$ and $K(\bm{f})$
 \begin{align}
 &\bar{w}^{i_1i_2}(\bm{f})=\sum_{\bm{k}\in J}(j2\pi\kappa)^{i_1+i_2}k_1^{i_1}k_2^{i_2}c_{k_1}c_{k_2}e^{j2\pi\bm{f}^T\bm{k}}\bm{b}(\bm{k}),\nonumber\\
  &w^{i_1i_2}(\bm{f})=\sum_{\bm{k}\in J}\delta_{{\Omega}^{c}}(\bm{k})(j2\pi\kappa)^{i_1+i_2}k_1^{i_1}k_2^{i_2}c_{k_1}c_{k_2}e^{j2\pi\bm{f}^T\bm{k}}\bm{b}(\bm{k}),
 \end{align}
 where $\delta_{{\Omega}^{c}}$ is iid Bernoulli random variable with parameter $p:=\frac{n^2-s}{n^2}$. We apply the result of the vector Bernstein inequality in Lemma \ref{vecbern} to the finite sequence of zero-mean random vectors of the form
  \begin{align}
 \bm{u}^{i_1i_2}(\bm{k})&:=(\delta_{\Omega^{c}}(\bm{k})-p)(j2\pi\kappa)^{i_1+i_2}k_1^{i_1}k_2^{i_2}c_{k_1}c_{k_2}e^{j2\pi\bm{f}^T\bm{k}}\bm{b}(\bm{k}),
 \end{align}  
 to demonstrate that the deviation between $w^{i_1i_2}(\bm{f})$ and the scaled version of $\bar{w}^{i_1i_2}(\bm{f})$ is small enough with high probability for $i_1,i_2 \in \{0,1,2,3\}$. We can write
   \begin{align}
w^{i_1i_2}(\bm{f})-p\bar{w}^{i_1i_2}(\bm{f})=\sum_{\bm{k}\in J}\bm{u}(\bm{k}).
 \end{align}
 To calculate $B$ in Lemma \ref{vecbern}, one can obtain an upper bound on $\ell_{2}$ norm of $\bm{u}$
   \begin{align}
\|\bm{u}(\bm{k})\|_2 &\le {\pi}^{i_1+i_2}\|\bm{c}\|^2_{\infty}\sup_{\bm{k}\in J}\|\bm{b}(\bm{k})\|_2\nonumber\\
&\le B:=\frac{7745}{m^2}\sqrt{r},
 \end{align}
 where the last inequality comes from (\ref{boundonkappa}) and (\ref{boundonc}), and $i_1=i_2=3$. Also, To compute 
 $\sigma^{2}$ in Lemma \ref{vecbern} we have
    \begin{align}
    &\sum_{\bm{k}\in J}\mathbb{E}\|\bm{u}^{i_1i_2}(\bm{k})\|^2_2=\sum_{\bm{k}\in J}c^2_{k_1}c^2_{k_2}\|\bm{b}(\bm{k})\|^2_2\nonumber\\&\cdot(2\pi\kappa)^{2i_1+2i_2} k_1^{2i_1}k_2^{2i_2}\mathbb{E}\big[(\delta_{\Omega^c}(\bm{l})-p)^2\big]\nonumber\\& \le 21r(2m+1)^2\pi^{2i_1+2i_2}\|\bm{c}\|^4_{\infty}\le\sigma^2:=\frac{240\times10^6r}{m^2}, \end{align} 
    where the first inequality is obtained from Lemma (\ref{boundonb}), (\ref{boundonkappa}), and the fact that  variance of Bernoulli model is equal to $p(1-p) \le 1$. The second inequality comes from (\ref{boundonc}) for $i_1=i_2=3$. By leveraging the result of the vector Bernstein inequality in Lemma \ref{vecbern}, we have 
       \begin{align}
       \label{eventw}
       & \mathbb{P}\Big[\sup_{\bm{f}\in \mathcal{G}}\|w^{i_1i_2}(\bm{f})-\bar{w}^{i_1i_2}(\bm{f})\|_{2}\geq t,\quad i_1,i_2 \in \{0,1,2,3\}\Big]\nonumber\\&\le 9|\mathcal{G}|\exp(\frac{-t^2}{8\sigma^2}+\frac{1}{4}),~~~~ \text{for}~0\le t \le \frac{\sigma^2}{B},
    \end{align}
    by union bound. The lower bound of probability is equal to $\epsilon/5$, if we set $t$ as 
        \begin{align}
      \label{equalt}
t:=\sigma\sqrt{8(\frac{1}{4}+\log\frac{45|\mathcal{G}|}{\epsilon})}.              \end{align}
          In the following, we show that this choice of $t$ satisfies $0\le t \le \frac{\sigma^2}{B}$,
                  \begin{align}
                \frac{t}{\sigma}=\sqrt{8(\frac{1}{4}+\log\frac{45|\mathcal{G}|}{\epsilon})}&\le\sqrt{86+32\log n+8\log\frac{1}{\epsilon}}\nonumber\\
                &\le0.4\sqrt{n}+\sqrt{8\log\frac{1}{\epsilon}},
        \end{align}
        where the last inequality comes from the fact that $\sqrt{86+32\log n}
        \le n$ for $n \geq 13$, consequently, $t\le \frac{\sigma^2}{B}$ if we set $C_r$ and $C_s$ small enough in Theorem \ref{thm.main}. The desired result is obtained for 
           \begin{align}
      \sqrt{\frac{768\times10^7r}{n^2}\big(\frac{1}{4}+\log\frac{36\times10^3n^4}{\epsilon}\big)} \le t\le C_v \big(\log(\frac{n^2}{\epsilon})\big)^{\frac{-1}{2}},
         \end{align}
        if we set $C_r$ small enough in Theorem \ref{thm.main}.
                \section{Proof of Lemma \ref{inverse result}}\label{proofboundebar}
                The proof involves the same techniques which were first proposed in \cite{tang2013compressed}. In the following Lemma, we demonstrate that the matrix $\bar{\bm{E}}$ is similar to the identity matrix and consequently is invertible.
       \begin{lem}\label{boundonEbar}(Proof in section \ref{prooflemmaboundonEbar}). If the conditions of Theorem  \ref{thm.main} hold, then 
       	\begin{align}
       		&\|\bm{I}-\bar{\bm{E}}\|\le 0.24,~~\|\bar{\bm{E}}\|\le 1.24,\\
       		&\|\bar{\bm{E}}^{-1}\|\le 1.32.
       	\end{align}
              \end{lem}
          It is possible to write $\bar{\bm{E}}$ and $\bm{E}$ in terms of the matrix $\bm{b}\bm{b}^{*}$. By
          the definition of $\bar{K}_{2D}$ and $K_{2D}$, we have 
         \begin{align}
         &\bar{\bm{E}}:=\sum_{\bm{k}\in J}c_{k_1}c_{k_2}\bm{b}(\bm{k})\bm{b}^*(\bm{k}),\\
         &\bm{E}:=\sum_{\bm{k}\in J}\delta_{\Omega^c}(\bm{k})c_{k_1}c_{k_2}\bm{b}(\bm{k})\bm{b}^*(\bm{k}),
         \end{align}
         One can show that $\bm{E}$ is concentrated around $\frac{n^2-s}{n^2}\bar{\bm{E}}$ with high probability.    
        We first define the self-adjoint zero mean matrix $\bm{X}$ as below
         \begin{align}
         \bm{X}(\bm{k}):=(p-\delta_{\Omega^c}(\bm{k}))c_{k_1}c_{k_2}\bm{b}(\bm{k})\bm{b}^*(\bm{k}),
         \end{align}
         where
         \begin{align}
         \hspace{-0.2cm}\mathbb{E}(\bm{X}(\bm{k}))=(p-\mathbb{E}(\delta_{\Omega^c}(\bm{k})))c_{k_1}c_{k_2}\bm{b}(\bm{k})\bm{b}^*(\bm{k})=0.
         \end{align}
         One can bound the the operator norm of $\bm{X}$ using Lemma \ref{boundonb} and the upper bound on the maximum of the vector $\bm{c}$ 
         \begin{align}
         \|\bm{X}(\bm{k})\|&\le \max_{\bm{k}\in J}\|c_{k_1}c_{k_2}\bm{b}(\bm{k})\bm{b}^{*}(\bm{k})\|\nonumber\\
         &\le\|\bm{c}\|^2_{\infty}\max_{\bm{k}\in J}\|\bm{b}(\bm{k})\|_2^2\le B:=\frac{36r}{m^2}.
         \end{align}
         Also, 
         \begin{align}
         \sum_{\bm{k}\in J}\mathbb{E}&\big(\bm{X}(\bm{k})\bm{X}^{T}(\bm{k})\big)=\nonumber\\&\bigg\|\sum_{\bm{k}\in J}c^2_{k_1}c^2_{k_2}\|\bm{b}(\bm{k})\|^2_2\bm{b}(\bm{k})\bm{b}^*(\bm{k})\mathbb{E}\big[(\delta_{\Omega^c}-p)^2\big]\bigg\|\nonumber\\
         &\le 21rp(1-p)\|\bm{c}\|^2_{\infty}\sum_{\bm{k}\in J}c_{k_1}c_{k_2}\bm{b}(\bm{k})\bm{b}^*(\bm{k})\nonumber\\
         &\le\frac{36rp}{m^2}\|\bar{\bm{E}}\|\le \sigma^2:=\frac{45rp}{m^2}.
         \end{align}
         where the first inequality uses the variance of the Bernoulli model with parameter $p$ and Lemma \ref{boundonb}, the second stems from  (\ref{boundonc}) and the definition of $\bar{\bm{E}}$ and the last one is the result of Lemma \ref{boundonEbar}. We set $t=\frac{p}{4}C_{\min}\big(\log\frac{n^2}{\epsilon}\big)^{- \frac{1}{2}}$ where $C_{\min}:=\min \{1,\frac{C_D}{4}\}$ for simplicity. By matrix  Bernstein inequality (Lemma \ref{matrixbernsitann}), we can write
         \begin{align}\label{eventd}
         &\mathbb{E}\Bigg\{\Big\|\bm{E}^{-1}-p\bar{\bm{E}}^{-1}\Big\|\geq t \Bigg\}\le \nonumber\\
         &6r\exp\Bigg(\frac{-pC^2_{\min}m^2}{32r}\bigg(45\log\frac{n^2}{\epsilon}+3C_{\min}\sqrt{\log\frac{n^2}{\epsilon}}\bigg)^{-1}\Bigg)\nonumber\\
         &\le6r\exp\Bigg(\frac{-C^{\prime}_D(n^2-s)}{r\log\frac{n^2}{\epsilon}}\Bigg),
         \end{align}
         for numerical constant $C^{\prime}_D$. The lower bound on this probability is $\epsilon/5$ under the conditions 
         \begin{align}
         &r \le \frac{C^{\prime}_{D}n^2}{2}\Bigg(\log\frac{30r}{\epsilon}\log\frac{n^2}{\epsilon}\Bigg)^{-1},~~s\le \frac{n^2}{2},
         \end{align}
         which hold under the assumption of Theorem \ref{thm.main}, if we set $C_r$ and $C_s$ small enough.
        
             Consequently, the lower bound on the smallest singular value of $\bm{E}$ can be obtained by triangle inequality,
     \begin{align}
    \hspace{-.3cm}\frac{\sigma_{\min}(\bm{E})}{p}\geq\sigma_{\min}(\bm{I})-\|\bm{I}-\bar{\bm{E}}\|-\frac{1}{p}\|\bm{E}-p\bar{\bm{E}}\|\geq 0.51.
     \end{align}
    Therefore, $\bm{E}$ is invertible.   \cite[Appendix E]{tang2013compressed} states that for any matrices $\bm{A}$ and $\bm{B}$ so that $\bm{B}$ is invertible and $\|\bm{A}-\bm{B}\|\|\bm{B}^{-1}\|\le \frac{1}{2}$ one can write
       \begin{align}
   &\|\bm{A}^{1}\|\le 2\|\bm{B}^{-1}\|,\nonumber\\
   &\|\bm{A}^{-1}-\bm{B}^{-1}\|\le 2\|\bm{B}^{-1}\|^2\|\bm{A}-\bm{B}\|.
     \end{align}
     Consider $\bm{A}:=\bm{E}$ and $\bm{B}:=p\bar{\bm{E}}$. Using Lemma (\ref{boundonEbar}) and (\ref{eventd}), we have
     \begin{align}
     \|\bm{E}-p\bar{\bm{E}}\|\big\|(p\bm{E})^{-1}\big\|\le \frac{1}{2},
     \end{align}
     with probability at least $1-\epsilon/5$. Based on  this and Lemma \ref{boundonEbar}, and event (\ref{eventd}) we also have
       \begin{align}
    &\|\bm{E}^{-1}\|\le 2\big\|(p\bar{\bm{E}})^{-1}\big\|\le \frac{4}{p},\nonumber\\
    &\big\|\bm{E}^{-1}-(p\bar{\bm{E}})^{-1}\big\|\le 2\big\|(p\bar{\bm{E}}^{-1})\big\|^2\|\bm{E}-p\bar{\bm{E}}\|\nonumber\\
    &\le\frac{C_D}{2p}\big(\log\frac{n^2}{\epsilon}\big)^{\frac{-1}{2}},
     \end{align}
     with the same probability. Regarding the conditions of Theorem \ref{thm.main} $s\le\frac{n^2}{2}$, therefore $\frac{1}{p}\le 2$. This concludes the proof.
     \section{Proof of Lemma \ref{boundonEbar} }\label{prooflemmaboundonEbar}
     We borrow some bound on the sub-matrices of $\bar{\bm{E}}$ from \cite{valiulahi2017two} under the minimum separation condition $\frac{1.68}{m}$.
     \begin{lem} (\cite[Lemma B.1]{valiulahi2017two}).\label{borrowedbund}
     	If the conditions of Theorem \ref{thm.main} hold, then 
     \begin{align}
   & \|\bm{I}-\bar{\bm{E}}_{00}\|_{\infty}\le3.17\times
   10^{-2},&& \|\kappa\bar{\bm{E}}_{10}\|_{\infty}\le4.35\times10^{-2},\nonumber\\
    & \|\kappa^2\bar{\bm{E}}_{11}\|_{\infty}\le4.6\times10^{-2},&& \|\bm{I}-\kappa^2\bar{\bm{E}}_{20}\|_{\infty}\le 0.15,
     \end{align}
    where the inequalities are obtained by  (\ref{boundonkappa}). Gershgorin's  disk theorem allows to bound the operator norm of the matrix $\bm{I}-\bar{\bm{E}}$
          \end{lem}
           \begin{align}
     \|\bm{I}-\bar{\bm{E}}\| &\le\|\bm{I}-\bar{\bm{E}}\|_{\infty }\nonumber\\ &\le\max\bigg\{\|\bm{I}-\bar{\bm{E}}_{00}\|_{\infty}+\|\kappa\bar{\bm{E}}_{01}\|_{\infty}+\|\kappa\bar{\bm{E}}_{10}\|_{\infty}\nonumber\\
      &,\|\kappa\bar{\bm{E}}_{10}\|_{\infty}+\|\kappa^2\bar{\bm{E}}_{11}\|_{\infty}+\|\bm{I}-\kappa^2\bar{\bm{E}}_{20}\|_{\infty}\bigg\} \le 0.24,
      \end{align}
      where the last inequality comes from  (\ref{boundonkappa}) and the results of the Lemma \ref{borrowedbund}. Consequently, 
      \begin{align}
            &\|\bar{\bm{E}}\|\le 1+\|\bm{I}-\bar{\bm{E}}\|_{\infty}\le 1.24,
      \end{align}
      and also,
      \begin{align}
          & \|\bar{\bm{E}}^{-1}\|\le \frac{1}{1-\|\bm{I}-\bm{E}\|_{\infty}}\le 1.32.
      \end{align}
                  \section{Proof of Lemma \ref{prop2}}\label{proofprop2}
                  It is possible to express $Q^{i_1i_2}(\bm{f})$ and  $\bar{Q}^{i_1i_2}(\bm{f})$ in terms of $\bm{h}$ and $\bm{r}$:
                   \begin{align}
      &\kappa^{i_1+i_2}\bar{Q}^{i_1i_2}(\bm{f})=\bar{w}^{i_1i_2}(\bm{f})^T\bar{\bm{E}}^{-1}\begin{bmatrix}
      \bm{h}\\
      \bm{0}\\ 
      \bm{0}
      \end{bmatrix},\\
       &\kappa^{i_1+i_2}Q(\bm{f})=w^{i_1i_2}(\bm{f})^T\bm{E}^{-1}\bigg(\begin{bmatrix}
         \bm{h}\\
         \bm{0}\\ 
         \bm{0}
         \end{bmatrix}-\frac{1}{\sqrt{n^2}}B_{\Omega}\bm{r}\bigg)\nonumber\\&\hspace{2cm}+\kappa^{i_1+i_2}R^{i_1i_2}(\bm{f}).
               \end{align}
               $Q^{i_1i_2}(\bm{f})$ and $\bar{Q}^{i_1i_2}(\bm{f})$ are related to each other as 
                   \begin{align}
        &\kappa^{i_1+i_2}Q^{i_1i_2}(\bm{f})=\kappa^{i_1+i_2}\bar{Q}^{i_1i_2}(\bm{f})+\kappa^{i_1+i_2}R^{i_1i_2}(\bm{f})\nonumber\\&+I_1^{i_1i_2}(\bm{f})+I_2^{i_1i_2}(\bm{f})+I_3^{i_1i_2}(\bm{f}),
        \end{align}
        in which
        \begin{align} &\hspace{-.44cm}I_1^{i_1i_2}(\bm{f}):=\frac{-1}{\sqrt{n^2}}w^{i_1i_2}(\bm{f})^T\bm{E}^{-1}B_{\Omega}\bm{r},\\
        &\hspace{-.44cm}I_2^{i_1i_2}(\bm{f}):=\bigg (w^{i_1i_2}(\bm{f})-\frac{n^2-s}{n^2}\bar{w}^{i_1i_2}(\bm{f})\bigg)^T\bm{E}^{-1}\begin{bmatrix}
        \bm{h}\\
        \bm{0}\\ 
        \bm{0}
        \end{bmatrix},\\
        &\hspace{-.44cm}I_3^{i_1i_2}(\bm{f}):=\frac{n^2-s}{n^2}\bar{w}^{i_1i_2}(\bm{f})^T\bigg(\bm{E}^{-1}-\frac{n^2}{n^2-s}\bar{\bm{E}}\bigg)\begin{bmatrix}
        \bm{h}\\
        \bm{0}\\ 
        \bm{0}
        \end{bmatrix}.
                \end{align}
                In the following Lemma, we show that there exists a bound on these terms on two-dimensional grid $\mathcal{G}$ with high probability.                 \begin{lem}(Proof in section \ref{proofboundondev}).\label{boundoni}
                	If the conditions of Theorem \ref{thm.main} hold, then the events
                	                 	\begin{align}
                	\varepsilon_{R}:=\bigg\{\sup_{\bm{f}\in \mathcal{G}}\big|\kappa^{i_1+i_2}R^{i_1i_2}(\bm{f})\big|\geq\frac{10^{-2}}{8},\quad i_1,i_2\in\{0,1,2,3\}\bigg\}, 
                	\end{align}
                	and
                \begin{align}
                                		\varepsilon_{i}:=\bigg\{\sup_{\bm{f}\in \mathcal{G}}\big|I_i^{i_1i_2}(\bm{f})\big|\geq\frac{10^{-2}}{8},\quad i_1,i_2\in\{0,1,2,3\}\bigg\},
                \end{align}
                for $i \in \{0,1,2,3\}$ and two-dimensional equispaced gird $\mathcal{G}$ with set size $800n^4$, happen with probability at most $\epsilon/5$ under the condition
                $\varepsilon^c_{B} \cap \varepsilon^c_{D}\cap \varepsilon^c_{v}$.
                
                Consequently, by triangle inequality we have
                \begin{align}
                \sup_{\bm{f}\in \mathcal{G}} \big|\kappa^{i_1+i_2}Q^{i_1i_2}(\bm{f})-\kappa^{i_1+i_2}\bar{Q}^{i_1i_2}(\bm{f})\big|\le \frac{10^{-2}}{2},
                              	                \end{align}\label{bounds}
                              	                with probability at least $1-\epsilon/5$ under the condition $\varepsilon^c_{B} \cap \varepsilon^c_{D}\cap \varepsilon^c_{v}$.
                \end{lem}
            We have already shown that the deviation between $Q^{i_1i_2}(\bm{f})$ and $\bar{Q}^{i_1i_2}(\bm{f})$ is small on a fine grid. Now we will extend this concept on the whole $[0,1]^2$.
            \begin{lem}(Proof in section \ref{proofQ}).\label{boundonQ}
            	If the conditions of Theorem \ref{thm.main} hold, then
            	\begin{align}
            		\kappa^{i_1+i_2}|Q^{i_1i_2}(\bm{f})-\bar{Q}^{i_1i_2}(\bm{f})|\le10^{-2},\quad i_1,i_2 \in \{0,1,2,3\}.
            	\end{align}
            \end{lem}
        We divide $[0,1]^2$ to two domains
        \begin{align}
&\mathcal{S}_{\text{near}}=\big\{\bm{f}|~\|\bm{f}-\bm{f}_i\|_{\infty}\le 0.09\big\},\nonumber\\&
\mathcal{S}_{\text{far}}=[0,1]^2\setminus\mathcal{S}_{\text{near}}.
        \end{align} 
        \cite{valiulahi2017two} has demonstrated that      $|\bar{Q}(\bm{f})|\le 0.9866$ for $\bm{f}\in \mathcal{S}_{\text{far}}$. One can leverage the result of Lemma \ref{boundonQ} and triangle inequality to obtain 
        \begin{align}
        |Q(\bm{f})|\le |\bar{Q}(\bm{f})|+10^{-2}\le 1,
        \end{align}
        for $\bm{f}\in \mathcal{S}_{\text{far}}$.\\
        Also, \cite{valiulahi2017two} has shown that the following Hessian matrix is negative definite in domain  $\bm{f}\in \mathcal{S}_{\text{near}}$, so $|\bar{Q}(\bm{f})|\le1$ in this domain,
        \begin{align}
                \bar{\bm{H}}=
        \begin{bmatrix}
        \bar{Q}^{20}(\bm{t})& \bar{Q}^{11}(\bm{t})\\
        \bar{Q}^{11}(\bm{t})& \bar{Q}^{02}(\bm{t})
        \end{bmatrix}.
        \end{align}
        More precisely, $\bar{Q}^{20}\le-1.4809m^2$, $\bar{Q}^{02}\le-1.4809m^2$ and  $|\bar{Q}^{11}|\le1.4743m^2$. It is possible to rewrite the elements of the matrix $\bar{\bm{H}}$ for $Q(\bm{f})$ then using the result of Lemma \ref{boundonQ}  
        \begin{align}\label{boundqs1}
        	Q^{20}(\bm{f})\le -1.5209m^2,\quad|Q^{11}(\bm{f})|\le 1.5143m^2,
        \end{align} 
        by (\ref{boundonkappa}). If the matrix $\bm{H}$ is concave, then $Q(\bm{f})<1$. The sufficient condition for concavity of this matrix is $\mathrm{Tr}(\bm{H})<0$ and $\mathrm{det}(\bm{H})>0$, where
          \begin{align}
        \label{eq49}
        &\text{Tr}(\bm{H})=Q^{20}(\bm{f})+Q^{02}(\bm{f}),\nonumber\\
        &\text{det}(\bm{H})=|Q^{20}(\bm{f})||Q^{02}(\bm{f})|-|Q^{11}(\bm{f})|^2.
        \end{align}
        By  (\ref{boundqs1}), it is easy to see that $\mathrm{Tr}(\bm{H})<0$ and $\mathrm{det}(\bm{H})>0$, so the Hessian matrix $\bm{H}$ is negative definite in the domain $\mathcal{S}_{near}$. This concludes the proof.
            \subsection{Proof of Lemma \ref{boundoni} }\label{proofboundondev}
            We follow a technique first proposed in \cite{fernandez2016demixing} to prove this lemma. 
            \begin{lem}(Hoeffding’s Inequality \cite{hoeffding1963probability}).\label{hoeffdin} If  the elements of  $\tilde{\bm{u}}$ are sampled  independently and identically distributed from a symmetric distribution on the complex unit circle, then for any $t$ and vector $\bm{u}$ one can write
            	\begin{align}
            	\mathbb{E}(|\langle\tilde{\bm{u}},\bm{u}\rangle|\geq t)\le 4\exp(\frac{-t^2}{4\|\bm{u}\|^2_2}). 
            	\end{align}
            	Consequently, the event 	
                        \begin{align}
            	\varepsilon=\bigg\{(|\langle\tilde{\bm{u}},\bm{u}\rangle|\geq \frac{10^{-2}}{8})\quad \forall\bm{u} \in \mathcal{U}\bigg\},
            	\end{align}
            	where $\mathcal{U}$ is a finite collection of vectors with size  $9|\mathcal{G}|=72\times10^2n^4$,
            	happens with probability
            	at most $\epsilon/20$ under the condition
            		\begin{align}
            \|\bm{u}\|_2^2 \le C_\mathcal{U}^2\bigg(\log\frac{n^2}{\epsilon}\bigg)^{-1},~~~C_\mathcal{U}=1/5000.
            	\end{align}
            	The proof follows from the union bound and the fact that $\big(\log\frac{576\times10^2n^4}{\epsilon}\big)^{-1}<\big(\log\frac{n^2}{\epsilon}\big)^{-1}$.
    
            	To bound $\mathbb{P}(\varepsilon_{R}|\varepsilon_{B}^c \cap\varepsilon_{E}^c \cap\varepsilon_{v}^c)$, one can use the following vector  
            	\begin{align}
            	\bm{u}^{i_1i_2}(\bm{f})&:=\frac{\kappa^{i_1+i_2}}{\sqrt{n^2}}\bigg[(j2\pi)^{i_1+i_2}k_{1_1}^{i_1}k_{1_2}^{i_2}e^{j2\pi\bm{f}^T\bm{k}_1}\cdots\nonumber\\
             &(j2\pi)^{i_1+i_2}k_{s_1}^{i_1}k_{s_2}^{i_2}e^{j2\pi\bm{f}^T\bm{k}_s}\bigg]^T,~~i_1,i_2 \in \{0,1,2,3\}
            	\end{align}
            \end{lem}
        in which $\bm{f}$ belongs to $\mathcal{G}$, so that $|\mathcal{U}|=9|\mathcal{G}|$ and also $\kappa^{i_1+i_2}R^{i_1i_2}(\bm{f})=\langle\bm{r},\bm{u}^{i_1i_2}(\bm{f})\rangle$. In the following, we show that $\|\bm{u}^{i_1i_2}(\bm{f})\|_2$ satisfy the criterion of Lemma \ref{hoeffdin}
        \begin{align}
       \|\bm{u}^{i_1i_2}(\bm{f})\|^2_2 &\le \frac{s( 2\pi m\kappa)^{2i_1+2i_2}}{n^2}\nonumber\\
        & \le\frac{s \pi^{12}}{n^2}\le C^2_\mathcal{U}\bigg(\log\frac{n^2}{\epsilon}\bigg)^{-1},
        \end{align}
        where the first inequality stems from the  union bound, the second one from (\ref{boundonkappa}), and the last one is obtained if we set $C_s$ small enough in the conditions of Theorem \ref{thm.main}.
        
        To bound $\mathbb{P}(\varepsilon_{1}|\varepsilon_{B}^c \cap\varepsilon_{E}^c \cap\varepsilon_{v}^c)$, we write
                \begin{align}
         I_1^{i_1i_2}(\bm{f})=\langle\bm{u}^{i_1i_2}(\bm{f}),\bm{r}\rangle
         \end{align}
         in which $\bm{u}^{i_1i_2}(\bm{f}):=\frac{-1}{\sqrt{n^2}}B^{*}_{\Omega}\bm{E}^{-1}w^{i_1i_2}(\bm{f})$ for $i_1,i_2\in \{0,1,2,3\}$, and also $\bm{f} \in \mathcal{G}$, so $|\mathcal{U}|=9|\mathcal{G}|$.
         
        One can obtain a bound on $\|\bm{u}^{i_1i_2}(\bm{f})\|_2$ using a bound on $\|w^{i_1i_2}(\bm{f})\|^2_2$ and Lemma \ref{event}. The following Lemma provides a bound on  $\ell_2$ norm of $w^{i_1i_2}(\bm{f})$
         \begin{lem}(Proof in section \ref{proofboundw}).\label{boundonnorm2w}
         	If the conditions of Theorem \ref{thm.main} hold, then 
         	\begin{align}
         	\|\bar{w}^{i_1i_2}(\bm{f})\|_2 \le C_{\bar{v}},
         	\end{align}
         where $C_{\bar{v}}$ is a fixed numerical constant.   Consequently,   
            \end{lem}           
         		\begin{align}
         	\|w^{i_1i_2}(\bm{f})\|_2 \le C_{\bar{v}}+C_{v},
         	\end{align}
         	 where we have used $\varepsilon_{v}^c$ in Lemma \ref{event} and triangle inequality and the facts that $\frac{n^2-s}{n^2}\le 1$ and $\big(\log\frac{n^2}{\epsilon}\big)^{-\frac{1}{2}}\le 1$.
         Combining this and the result of Lemma \ref{eventd},
        \begin{align}\label{CBdefine}
        \|\bm{u}^{i_1i_2}(\bm{f})\|_2 &\le \frac{1}{\sqrt{n^2}}\|\bm{B}_{\Omega}\|\|\bm{E}^{-1}\|\|w^{i_1i_2}(\bm{f})\|_2\nonumber\\
        &\le\frac{8(C_{v}+C_{\bar{v}})\|\bm{B}_{\Omega}\|}{\sqrt{n^2}}
               \end{align}
            under the condition $\varepsilon^c_{D}\cap \varepsilon^c_{v}$. If 
            \begin{align}\label{defcb}
            \|\bm{B}_{\Omega}\|\le C_B \big(\log\frac{n^2}{\epsilon}\big)^{\frac{-1}{2}}\sqrt{n^2}, \quad C_B:=\frac{C_v}{8(C_{v}+C_{\bar{v}})},
            \end{align}
             then the desired bound in Lemma \ref{hoeffdin} is obtained for numerical constant $C_{\mathcal{U}}$. The condition on $\|\bm{B}_{\Omega}\|$ is satisfied by $\varepsilon_{B}^{c}$ in Lemma \ref{boundbomega}.
            
           To bound $\mathbb{P}(\varepsilon_{2}|\varepsilon_{B}^c \cap\varepsilon_{E}^c \cap\varepsilon_{v}^c)$, we consider
            \begin{align}
            &I^{i_1i_2}(\bm{f})=\langle\bm{u}^{i_1i_2}(\bm{f}),\bm{h}\rangle\nonumber\\
            &\bm{u}^{i_1i_2}(\bm{f}):=\bm{P}\bm{E}^{-1}\big(w^{i_1i_2}(\bm{f})-\frac{n^2-s}{n^2}\bar{w}^{i_1i_2}(\bm
            {f})\big),
            \end{align}
            where $\bm{P}\in\mathbb{R}^{r\times 3r}$ is a projection matrix that takes the first $r$ entries in a vector. One can obtain an upper bound on $\ell_{2}$ norm of $\bm{u}^{i_1i_2}(\bm{f})$ using Lemma \ref{eventd} in $\varepsilon^c_{D}$ as follow
            \begin{align}
            \|\bm{u}^{i_1i_2}(\bm{f})\|_2&\le \|\bm{P}\|\|\bm{E}^{-1}\|\bigg\|w^{i_1i_2}(\bm{f})-\frac{n^2-s}{n^2}\bar{w}^{i_1i_2}(\bm{f})\bigg\|_2\nonumber\\&\le8\bigg\|w^{i_1i_2}(\bm{f})-\frac{n^2-s}{n^2}\bar{w}^{i_1i_2}(\bm{f})\bigg\|_2,
            \end{align}
            where the last inequality comes from the fact that $\|\bm{P}\|=1$ and it is valid for $\bm{f} \in \mathcal{G}$, so that $|\mathcal{U}|=9|\mathcal{G}|$. If
                   \begin{align}\label{defccbb}
          	&\bigg\|w^{i_1i_2}(\bm{f})-\frac{n^2-s}{n^2}\bar{w}^{i_1i_2}(\bm{f})\bigg\|_2\le C_v\bigg(\log\frac{n^2}{\epsilon}\bigg)^{\frac{-1}{2}},\nonumber\\&C_v:=\frac{C_\mathcal{U}}{8}.
          	          \end{align}
          	          Then the desired bound in Lemma \ref{hoeffdin} is achieved for the numerical constant $C_{v}$. The condition on the deviation of $w^{i_1i_2}(\bm{f})$ and  $\frac{n^2-s}{n^2}w^{i_1i_2}(\bm{f})$ holds with respected to $\varepsilon^c_{v}$ in Lemma \ref{event}.
          	                   	          
          	          Finally, we bound $\mathbb{P}(\varepsilon_{3}|\varepsilon_{B}^c \cap\varepsilon_{E}^c \cap\varepsilon_{v}^c)$. Let
          	          \begin{align}
          	          &I_3^{i_1i_2}(\bm{f})=\langle\bm{u}^{i_1i_2}(\bm{f}),\bm{h}\rangle,\nonumber\\&
          	          \bm{u}^{i_1i_2}(\bm{f}):=\frac{n^2-s}{n^2}\bm{P}\bigg(\bm{E}^{-1}-\frac{n^2}{n^2-s}\bar{\bm{E}}^{-1}\bigg)\bar{w}^{i_1i_2}(\bm{f}).
          	          \end{align}
          	          One can  bound  $\ell_{2}$ norm of $\bm{u}^{i_1i_2}(\bm{f})~\forall \bm{f} \in \mathcal{G}$ for $i_1,i_2\in \{0,1,2,3\}$, so $|\mathcal{U}|=9|\mathcal{G}|$
          	                     	         \begin{align}
          	        \|\bm{u}^{i_1i_2}(\bm{f})\|_2&\le \|\bm{P}\|\big\|\bm{E}^{-1}-\frac{n^2}{n^2-s}\bar{\bm{E}}^{-1}\big\|\|\bar{w}^{i_1i_2}(\bm{f})\|_2\nonumber\\&\le C_{\bar{v}}\big\|\bm{E}^{-1}-\frac{n^2}{n^2-s}\bar{\bm{E}}^{-1}\big\|,
          	         \end{align} 
          	         where the last inequality comes from Lemma \ref{boundonnorm2w} and $\|\bm{P}\|=1$. If
          	                   	         \begin{align}\label{definecd}
          	         \big\|\bm{E}^{-1}-\frac{n^2}{n^2-s}\bar{\bm{E}}^{-1}\big\|\le C_D 
          	         \bigg(\log\frac{n^2}{\epsilon}\bigg)^{\frac{-1}{2}},\quad C_D:=\frac{C_{\mathcal{U}}}{C_{\bar{v}}},
          	          \end{align}         	                   	          then the desired bound in Lemma \ref{hoeffdin} is obtained for the numerical constant $C_D$. The condition on the deviation of $\bm{E}$ and $\frac{n^2}{n^2-s}\bar{\bm{E}}^{-1}$ is satisfied by  $\varepsilon^c_{D}$ in Lemma \ref{eventd}.
          \subsection{Proof of Lemma \ref{boundonnorm2w}}	\label{proofboundw} 
          Regarding the fact that $\|\cdot\|_2 \le \|\cdot\|_{1} $, we have
                    \begin{align}
          	&\|\bar{w}^{i_1i_2}(\bm{f})\|_2\le \|\bar{w}^{i_1i_2}(\bm{f})\|_1\nonumber\\
          	&=\sum_{i=1}^{k}\kappa^{i_1+i_2}|\bar{K}^{i_1i_2}(\bm{f}-\bm{f}_i)|+\sum_{i=1}^{k}\kappa^{i_1+i_2+1}|\bar{K}^{i_1+1,i_2}(\bm{f}-\bm{f}_i)|\nonumber\\&\sum_{i=1}^{k}\kappa^{i_1+i_2+1}|\bar{K}^{i_1,i_2+1}(\bm{f}-\bm{f}_i)|.
          	          	          \end{align}  
          	          	          We first borrow the result of Lemma H.10 from \cite{fernandez2016super} 	          to bound the magnitude of two-dimensional kernel in unit square          	          	 \begin{lem}(\cite[Lemma H.10]{fernandez2016super})\label{lemmaboundonk1}
      	\begin{align}
      	\kappa^{\ell}|\bar{K}^{\ell}(f)|\le 
      	\bigg \{
      	\begin{array}{rl}
      	C_1\quad\quad\quad\quad&|f|\le \frac{80}{m},\\
      C_2m^{-3}|f|^{-3}&  \frac{80}{m}\le |f|\le \frac{1}{2},
      	\end{array}
      	\end{align}
      	where $C_1$ and $C_2$ are numerical constants. 
         \end{lem}
    Since $\bar{K}(f_1,f_2)=\bar{k}(f_1)\bar{k}(f_2)$, one can extend the mentioned Lemma to two-dimensional case. Without loss of generality, let us map the unit square to $[-\frac{1}{2},\frac{1}{2}]^2$, so we have
     \begin{align}
     \kappa^{i_1+i_2}|\bar{K}^{i_1,i_2}(\bm{f})|\le 
     \Bigg \{
     \begin{array}{rl}
     C_1^2\quad\quad\quad\quad\quad\quad\quad& a:\|\bm{f}\|_\infty\le \frac{80}{m},\\
     C_2^2m^{-6}|f_1|^{-3}|f_2|^{-3}&  [-\frac{1}{2},\frac{1}{2}]^2\setminus a
     \end{array}
          \end{align}
          for $i_1,i_2 \in \{0,1,2,3\}$, where $C_1$ and $C_2$ are numerical constants defined in Lemma \ref{lemmaboundonk1}.\\
   The minimum separation condition implies that there exist at most $95^2$  support elements  in the square $\|\bm{f}-\bm{f}_i\|_{\infty}\le \frac{80}{m}$. We bound those elements by $C_{1}^2$ and use the decreasing bound to handle the remaining components
  \begin{align}
  	&\hspace{-0.5cm}\sum_{i=1}^{k}\kappa^{i_1+i_2}|\bar{K}^{i_1i_2}(\bm{f}-\bm{f}_i)|\le\hspace{-0.5cm} \sum_{b:\|\bm{f}-\bm{f}_i\|_{\infty}\le \frac{80}{m}}\hspace{-0.5cm}C_1^2\nonumber\\
  	&\hspace{-0.5cm}+\sum_{[-\frac{1}{2},\frac{1}{2}]^2\setminus b}\frac{C_2^2}{m^6|f_1-f_{1i}|^3|f_1-f_{1i}|^3}\le 95^2C_1^2\nonumber\\&\hspace{-0.5cm}+4C_2^2\bigg(\sum_{i=1}^{\infty}\frac{1}{m^3(i\Delta_{\min})^3}\bigg)^2\le 95^2C_1^2+0.18C_2^2\zeta^2(3),
  	  \end{align}
  	  where $\zeta(3):=\sum_{i=1}^{\infty}\frac{1}{i^3}$ is known as Apery's constant which can be bounded by $1.21$. Consider $C_{\bar{v}}=95^2C_{1}^{2}+0.27C_{2}^2$, so the proof is complete.
  	  \subsection{Proof of Lemma \ref{boundonQ}}\label{proofQ}
         	  We first use the bound on the deviation between $Q^{i}(f)$ and  $\bar{Q}^{i}(f)$ from \cite{fernandez2016super}
         	 \begin{align}
         	 \kappa^i|Q^{i}(f_1)-Q^{i}(f_2)|\le n^2 |f_1-f_2|,\nonumber\\
         	 \kappa^i|\bar{Q}^{i}(f_1)-\bar{Q}^{i}(f_2)|\le n^2 |f_1-f_2|,
         	 \end{align} 
         	 for any $f_1,f_2 \in [0,1]$ and $i \in \{0,1,2,3\}$. Also, consider another useful bound $\kappa^{i}|\bar{K}(f)|\le Cn^2$ where $C$ is a numerical constant. One can extend this concept to two-dimensional case using Bernstein's polynomial inequality as below,  
         	 \begin{align}\label{boundss}
         	 \kappa^{i_1+i_2}|&Q(\bm{f}_1)-Q(\bm{f}_2)|\le \kappa^{i_1+i_2}\big(|Q(\bm{f}_1)-Q(f_{11},f_{22})|\nonumber\\
         	 &+|Q(f_{11},f_{22})-Q(\bm{f}_2)|\big)\le C^2n^4|f_{12}-f_{22}|\nonumber\\&+C^2n^4|f_{11}-f_{21}|\le 2C^2n^4\|\bm{f}_1-\bm{f}_2\|_{\infty}.
         	 \end{align}
         	 \cite{chi2015compressive} has done a similar analysis in the proof of Lemma 7. For any $\bm{f} \in [0,1]^2$ there exists two-dimensional grid point $\bm{f}_{\mathcal{G}}$ such that the maximum distances between two points is smaller than the step size $(800C^2n^4)^{-1}$. So, Lemma \ref{bounds} and triangle inequality lead to          	  \begin{align}
         	  	\kappa^{i_1+i_2}&\big|Q^{i_1i_2}(\bm{f})-\bar{Q}^{i_1i_2}(\bm{f})\big|\nonumber\\&\le \kappa^{i_1+i_2}\big|Q^{i_1i_2}(\bm{f})-Q^{i_1i_2}(\bm{f}_{\mathcal{G}})\big|\nonumber\\&+\kappa^{i_1+i_2}\big|Q^{i_1i_2}(\bm{f}_{\mathcal{G}})-\bar{Q}^{i_1i_2}(\bm{f}_{\mathcal{G}})\big|\nonumber\\&+\kappa^{i_1+i_2}\big|\bar{Q}^{i_1i_2}(\bm{f}_{\mathcal{G}})-\bar{Q}^{i_1i_2}(\bm{f})\big|\nonumber\\& \le 4C^2n^4\|\bm{f}-\bm{f}_{\mathcal{G}}\|_{\infty}+5 \times10^{-3}\le 10^{-2}.
         	  \end{align} 
         	  And the proof is complete. 
         	  \subsection{Proof of Lemma \ref{prop3}}  \label{ProoflemmaC}
         	   One can recast the coefficient $C_{\bm{k}}$ in terms of $\bm{h}$ and $\bm{r}$. Let $\bm{k}$ be an arbitrary element of $\Omega^c$
         	  \begin{align}
         	  &C_{\bm{k}}=c_{k_1}c_{k_2}\Bigg(\sum_{i=1}^{r}\alpha_ie^{j2\pi\bm{f}^T_i\bm{k}}\nonumber\\&+i2\pi\kappa k_1\sum_{i=1}^{r}\beta_{1i}e^{i2\pi\bm{f}_{i}^{T}\bm{k}}+i2\pi\kappa k_2\sum_{i=1}^{r}\beta_{2i}e^{i2\pi\bm{f}_{i}^{T}\bm{k}}\Bigg)\nonumber\\&=c_{k_1}c_{k_2}\bm{b}(\bm{k})^{*}\begin{bmatrix}
         	  \bm{\alpha}\\
         	  \bm{\beta}_1\\ 
         	  \bm{\beta}_2
         	  \end{bmatrix}
         	  =c_{k_1}c_{k_2}\bm{b}(\bm{k})^{*}E^{-1}\bigg(\begin{bmatrix}
         	  \bm{h}\\
         	  \bm{0}\\ 
         	  \bm{0}
         	  \end{bmatrix}-\frac{1}{\sqrt{n^2}}B_{\Omega}\bm{r}\bigg)\nonumber\\&=c_{k_1}c_{k_2}\bigg(\langle\bm{p}\bm{E}^{-1}\bm{b}(\bm{k}),\bm{h}\rangle+\frac{1}{\sqrt{n^2}}\bm\langle{B}^{*}_{\Omega}\bm{E}^{-1}\bm{b}(\bm{k}),\bm{r}\rangle\bigg),
         	  \end{align}
         	   where $\bm{P}\in\mathbb{R}^{r\times3r}$ is the projection matrix that retains the first $r$ entries of a vector.\\
         	   To bound $|C_{\bm{k}}|$, we first obtain a bound on  $\bm{P}\bm{E}^{-1}\bm{b}$ as below
         	   \begin{align}
         	   	\|\bm{P}\bm{E}^{-1}\bm{b}(\bm{k})\|^2_2&\le\|\bm{P}\|^2\|\bm{E}^{-1}\|^2\|\bm{b}(\bm{k})\|^2_2\nonumber\\
         	   	&\le 1344r\le \frac{0.07^2n^2}{\log\frac{40}{\epsilon}},
         	   \end{align}
         	   where the last inequality is obtained under the conditions of Theorem \ref{thm.main} when we set $C_k$ small enough and the second one is a combination of Lemmas \ref{boundonb} , \ref{inverse result} and the fact that $\|\bm{P}\|_{2}=1$. Also, we have
         	   \begin{align}
         	   \|\bm{B}^{*}_{\Omega}\bm{E}^{-1}\bm{b}(\bm{k})\|_2^2&\le \|\bm{B}_{\Omega}\|^2\|\bm{E}^{-1}\|^2\|\bm{b}(\bm{k})\|^2_2\nonumber\\& \le1344rn^2C^2_B\le \frac{0.07^2n^2}{\log\frac{40}{\epsilon}},
         	   \end{align}
         	   where the second inequality stems from Lemmas \ref{boundonb} and \ref{boundbomega} and the last one comes from the assumption of Theorem \ref{thm.main} if we set $C_r$ small enough.\\
         	   One can obtain $\epsilon/10$ for the minimum probability of the following events
         	   by Hoeffding's inequality 
         	   \begin{align}
         	   &|\langle\bm{P}\bm{E}^{-1}\bm{b}(\bm{k}),\bm{h}\rangle|> 0.07\sqrt{n^2},\nonumber\\
         	   &|\langle\bm{B}^{*}_{\Omega}\bm{E}^{-1}\bm{b}(\bm{k}),\bm{r}\rangle|> 0.07n^2.
         	            	   \end{align}
         	            	   Using $\|\bm{c}\|_{\infty}\le \frac{1.3}{m}$ and the union bound, we have
         	            	   \begin{align}
         	            	  |C_{\bm{k}}|\le \frac{2.6^2}{n^2}\big(0.07\sqrt{n^2}+0.07\sqrt{n^2}\big)\le \frac{1}{\sqrt{n^2}},
         	            	   \end{align}
         	            	   with probability at least $1-\epsilon/5$. This concludes the proof.
         	            	   
         	            	   \subsection{Proof of Lemma \ref{duall}}\label{proofdu}
         	            	   Consider $\bm{C}\in \mathbb{C}^{n \times n}$ as the dual variable. By Lagrangian theorem we have 
         	            	   
         	            	   \begin{align}
         	            	   &\mathcal{L}(\tilde{\bm{\mu}},\tilde{\bm{Z}},\bm{C})=\|\tilde{\bm{\mu}}\|_{\mathrm{TV}}+\lambda\|\tilde{\bm{Z}}\|_{1}+\langle\bm{Y}-\mathcal{F}\tilde{\bm{\mu}}-\tilde{\bm{Z}},\bm{C}\rangle_{F}\nonumber\\
         	            	   &=\|\tilde{\bm{\mu}}\|_{\mathrm{TV}}\bigg(1-\frac{\langle\bm{\mathcal{F}}^{*}\bm{C},\tilde{\bm{\mu}}\rangle}{\|\tilde{\bm{\mu}}\|_{\mathrm{TV}}}\bigg)+\lambda \|\bm{Z}\|_{1}\bigg(1-\frac{\langle\bm{C},\bm{Z}\rangle}{\lambda\|\tilde{\bm{Z}}\|_{1}}\bigg)\nonumber\\
         	            	   &+\langle\bm{C},\bm{Y}\rangle.
         	            	   \end{align}
         	            	   One can minimize this function with respect to $\tilde{\bm{\mu}}$ and $\tilde{\bm{Z}}$ simultaneously, then maximize over dual variable $\bm{C}$ \cite{boyd2004convex}. At the first. by Holder's inequality, we have
         	            	   \begin{align}
         	            	   \frac{\langle\bm{\mathcal{F}}^{*}\bm{C},\tilde{\bm{\mu}}\rangle}{\|\tilde{\bm{\mu}}\|_{\mathrm{TV}}}\le \|\bm{\mathcal{F}}^{*}\bm{C}\|_{\infty}, ~~~~\frac{\langle\bm{C},\bm{Z}\rangle}{\|\tilde{\bm{Z}}\|_{1}}\le \|\bm{C}\|_{\infty}.
         	            	   \end{align}
         	            	   If $\|\bm{\mathcal{F}}^{*}\bm{C}\|_{\infty} \le 1$ and $\|\bm{C}\|_{\infty}\le \lambda$, then the minimum occurs at $\langle\bm{C},\bm{Y}\rangle$ otherwise, the problem is unbounded below. For second, we should maximize this term on $\bm{C}$ as below
         	            	   \begin{align}
         	            	   \max_{\bm{C}\in \mathbb{C}^{n \times n}}  \left\{ \begin{array}{ll}
         	            	   \langle\bm{C},\bm{Y}\rangle & \mbox{\text{if}~~ $\|\bm{\mathcal{F}}^{*}\bm{C}\|_{\infty} \le 1,~~\|\bm{C}\|_{\infty}\le \lambda $};\\
         	            	   -\infty & \mbox{\text{otherwise}}.\end{array} \right. 
         	            	   \end{align}
         	            	   Converting the implicit conditions into explicit conditions concludes the proof.
         \bibliographystyle{ieeetr}
         \bibliography{HBReference}
\end{document}